\documentclass[preprint, preprintnumbers ,amsmath,amssymb]{revtex4}
\usepackage{amssymb}
\usepackage{amsmath}
\usepackage{graphicx}
\usepackage{multirow}

\begin{document}



\title{New Perspectives on the Schr{\"o}dinger-Pauli Theory of Electrons:
Part II: Application to the Triplet State of a Quantum Dot in a
Magnetic Field}



\author{Marlina Slamet$^{1}$ and Viraht Sahni$^{2}$}

\affiliation{$^{1}$Sacred Heart University, Fairfield, Connecticut
06825
\\
$^{2}$Brooklyn College and The Graduate School of the City
University of New York, New York, New York 10016.}


\date{\today}

\begin{abstract}
The Schr{\"o}dinger-Pauli theory of electrons in the presence of a
static electromagnetic field can be described from the perspective
of the individual electron via its equation of motion or `Quantal
Newtonian' first law. The law is in terms of `classical' fields
whose sources are quantum-mechanical expectation values of Hermitian
operators taken with respect to the wave function.  The law states
that the sum of the external and internal fields experienced by each
electron vanishes. The external field is the sum of the binding
electrostatic and Lorentz fields.  The internal field is the sum of
fields representative of properties of the system: electron
correlations due to the Pauli exclusion principle and Coulomb
repulsion; the electron density; kinetic effects; the current
density.  Thus, the internal field is a sum of the
electron-interaction, differential density, kinetic, and internal
magnetic fields.  The energy can be expressed in integral virial
form in terms of these fields.  Via this perspective, the
Schr{\"o}dinger-Pauli equation can be written in a generalized form
which then shows it to be intrinsically self-consistent.  This new
perspective is explicated by application to the triplet $2^{3}S$
state of a $2$-D $2$-electron quantum dot in a magnetic field. The
quantal sources of the density; the paramagnetic, diamagnetic, and
magnetization current densities; pair-correlation density; the
Fermi-Coulomb hole charge; and the single-particle density matrix
are obtained, and from them the corresponding fields determined. The
fields are shown to satisfy the `Quantal Newtonian' first law. The
components of the energy too are determined from these fields.
Finally, the example is employed to demonstrate the intrinsic
self-consistent nature of the Schr{\"o}dinger-Pauli equation.
\end{abstract}

\pacs{}

\maketitle



\section{Introduction}

The Schr{\"o}dinger-Pauli theory \cite{1} is a description of a
system of $N$ electrons in the presence of an external electrostatic
binding field ${\boldsymbol{\cal{E}}} ({\bf{r}}) = -
{\boldsymbol{\nabla}} v ({\bf{r}})/e$ and a magnetostatic field
${\boldsymbol{\cal{B}}} ({\bf{r}}) = {\boldsymbol{\nabla}} \times
{\bf{A}} ({\bf{r}})$, where $v ({\bf{r}})$ and ${\bf{A}} ({\bf{r}})$
are scalar and vector potentials. In the theory the interaction of
the magnetic field with both the orbital and spin angular momentum
is explicitly considered. The stationary-state Schr{\"o}dinger-Pauli
differential equation is (charge of electron $- e$)
\begin{equation}
\bigg[ \frac{1}{2 m} \sum_{k} \big(\hat{\bf{p}}_{k} + \frac{e}{c}
{\bf{A}} ({\bf{r}}_{k}) \big)^{2} + g \mu_{B} \sum_{k}
{\boldsymbol{\cal{B}}} ({\bf{r}}_{k}) \cdot {\bf{s}}_{k} + \hat{W} +
\hat{V} \bigg] \Psi ({\bf{X}}) = E \Psi ({\bf{X}}),
\end{equation}
where the canonical momentum operator $\hat{\bf{p}} = - i \hbar
{\boldsymbol{\nabla}}$, the gyromagnetic ratio is $g$, the Bohr
magneton $\mu_{B} = e\hbar/2mc$, the velocity of light is $c$, the
spin angular momentum vector is ${\bf{s}}$, the electron interaction
operator $\hat{W} = \frac{1}{2} \sideset{}{'}\sum_{k,\ell}
e^{2}/|{\bf{r}}_{k} - {\bf{r}}_{\ell}|$ , the external binding
potential operator $\hat{V} = \sum_{k} v ({\bf{r}}_{k})$, the wave
function is $\Psi ({\bf{X}})$, the eigenenergy is $E$, and ${\bf{X}}
= {\bf{x}}_{1}, \ldots, {\bf{x}}_{N}$, with ${\bf{x}} = {\bf{r}}
\sigma$, the spatial and spin coordinates, respectively.

In the previous paper \cite{2}, referred to as Part I, the
Schr{\"o}dinger-Pauli theory of electrons was described from a new
perspective.  In the present work, we explicate the new perspective
by application to the triplet $2^{3} S$ state of a $2$-electron
$2$-dimensional `artificial atom' or quantum dot in a magnetic field
\cite{3,4,5,6}.  The motion of the electrons of the `artificial
atom' is confined to $2$ dimensions in a quantum well in a thin
layer of a semiconductor such as GaAs which is sandwiched between
two layers of another semiconductor AlGaAs.  The $2$-dimensional
motion of the electrons is restricted by an electrostatic field that
can be varied.  This motion can be further constrained by a magnetic
field perpendicular to the plane of motion.   As the `artificial
atom' is in a semiconductor, the free electron mass m of Eq. (1)
must be replaced by the band effective mass $m^{\star}$, and the
electron-interaction modified by the dielectric constant $\epsilon$.
For GaAs the effective mass is $m^{\star} = 0.067 m$, and $\epsilon
= 12.4$. Finally, the binding potential $v ({\bf{r}})$ of the
electrons in a quantum dot has been established via both theory and
experiment to be harmonic \cite{5,6,7}.  In spite of the reduced
dimensionality, and the fact that the size of a quantum dot is an
order of magnitude greater than that of a natural atom, such
`artificial atoms' exhibit very similar electronic structure.  The
stationary-state Schr{\"o}dinger-Pauli Hamiltonian for an $N$
electron quantum dot in a magnetic field ${\boldsymbol{\cal{B}}}
({\bf{r}})$ is thus the following:
\begin{eqnarray}
\bigg[ \frac{1}{2 m^{\star}} \sum_{k} \big(\hat{\bf{p}}_{k} +
\frac{e}{c} {\bf{A}} ({\bf{r}}_{k}) \big)^{2} &+& g^{\star} \mu_{B}
\sum_{k} {\boldsymbol{\cal{B}}} ({\bf{r}}_{k}) \cdot {\bf{s}}_{k} +
\frac{1}{2 \epsilon} \sideset{}{'}\sum_{k,\ell} \frac{e^{2}}
{|{\bf{r}}_{k} - {\bf{r}}_{\ell}|}  \nonumber\\
&+& \frac{1}{2 m^{\star}} \sum_{k} \omega^{2}_{0} r^{2}_{k} \bigg]
\Psi ({\bf{X}}) = E \Psi ({\bf{X}}),
\end{eqnarray}
where $g^{\star}$ is the corresponding gyromagnetic ratio, and
$\omega_{0}$ the binding harmonic frequency.

The new description of stationary-state Schr{\"o}dinger-Pauli theory
is from the perspective of the individual electron via its equation
of motion or `Quantal Newtonian' first law.  For details of this
perspective we refer the reader to Part I.  However, for an
understanding of the present paper independent of Part I, we provide
a brief review of the new perspective.

According to the law, each electron experiences an external
${\boldsymbol{\cal{F}}}^\mathrm{ext} ({\bf{r}})$ and an internal
${\boldsymbol{\cal{F}}}^\mathrm{int} ({\bf{r}})$ field, the sum of
which vanishes:
\begin{equation}
{\boldsymbol{\cal{F}}}^\mathrm{ext} ({\bf{r}}) +
{\boldsymbol{\cal{F}}}^\mathrm{int} ({\bf{r}}) = 0.
\end{equation}
The external field is the sum of the electrostatic
${\boldsymbol{\cal{E}}} ({\bf{r}})$ and Lorentz
${\boldsymbol{\cal{L}}} ({\bf{r}})$ fields:
\begin{equation}
{\boldsymbol{\cal{F}}}^\mathrm{ext} ({\bf{r}}) =
{\boldsymbol{\cal{E}}} ({\bf{r}}) - {\boldsymbol{\cal{L}}}
({\bf{r}}).
\end{equation}
The internal field is the sum of the electron-interaction
${\boldsymbol{\cal{E}}}_{ee} ({\bf{r}})$, kinetic
${\boldsymbol{\cal{Z}}} ({\bf{r}})$, differential density
${\boldsymbol{\cal{D}}} ({\bf{r}})$, and an internal magnetic
${\boldsymbol{\cal{I}}}_{m} ({\bf{r}})$ field component:
\begin{equation}
{\boldsymbol{\cal{F}}}^\mathrm{int} ({\bf{r}}) =
{\boldsymbol{\cal{E}}}_{ee} ({\bf{r}}) - {\boldsymbol{\cal{Z}}}
({\bf{r}}) - {\boldsymbol{\cal{D}}} ({\bf{r}}) -
{\boldsymbol{\cal{I}}}_{m} ({\bf{r}}).
\end{equation}

These fields are, respectively, representative of electron
correlations due to the Pauli exclusion principle and Coulomb
repulsion, the kinetic effects, the electron density, and the
physical current density.  The sources of these fields and the
Lorentz field, are quantum-mechanical expectation values of
Hermitian operators taken with respect to the wave function $\Psi$.
The individual fields are not necessarily conservative.  However,
the sum $\{ {\boldsymbol{\cal{E}}}_{ee} ({\bf{r}}) -
{\boldsymbol{\cal{Z}}} ({\bf{r}}) - {\boldsymbol{\cal{D}}}
({\bf{r}}) - {\boldsymbol{\cal{I}}}_{m} ({\bf{r}}) -
{\boldsymbol{\cal{L}}} ({\bf{r}}) \}$ is always conservative. The
`Quantal Newtonian' first law is valid for \emph{arbitrary} state.
The definitions of the various quantal sources and of their
respective fields will be provided as each property of the triplet
state of the quantum dot is discussed.

In Sect. II we discuss the structure and properties of the
closed-form analytical complex wave function $\Psi ({\bf{X}})$ for
the triplet $2^{3} S$ state of a two-electron quantum dot.  In
particular, the nodal structure of the wave function, and the
satisfaction by the wave function of the integral nodal
electron-electron coalescence condition \cite{8,9,10,11}. Further,
we describe the parity of the wave function about the various nodes,
particularly about the origin and points of electron-electron
coalescence.   For the derivation of the wave function we follow the
method of Taut \cite{12,13,14,15,16,17,18}. The \emph{local} quantal
sources of the electronic density $\rho ({\bf{r}})$, and physical
current density ${\bf{j}} ({\bf{r}})$ together with its paramagnetic
${\bf{j}}_{p} ({\bf{r}})$, diamagnetic ${\bf{j}}_{d} ({\bf{r}})$,
and magnetization ${\bf{j}}_{m} ({\bf{r}})$ density components, and
the \emph{nonlocal} sources of the single-particle density matrix
$\gamma ({\bf{r r}}')$, the pair-correlation density $g ({\bf{r
r}}')$, and the Fermi-Coulomb hole charge $\rho_{xc} ({\bf{r r}}')$,
are described in Sect. III. The various fields that arise from these
quantal sources are discussed in Sect. IV. These fields comprise the
electron-interaction ${\boldsymbol{\cal{E}}}_{ee} ({\bf{r}})$ and
its Hartree ${\boldsymbol{\cal{E}}}_{H} ({\bf{r}})$ and
Pauli-Coulomb ${\boldsymbol{\cal{E}}}_{xc} ({\bf{r}})$ components;
the kinetic ${\boldsymbol{\cal{Z}}} ({\bf{r}})$; the differential
density ${\boldsymbol{\cal{D}}} ({\bf{r}})$; internal magnetic
${\boldsymbol{\cal{I}}}_{m} ({\bf{r}})$; and the Lorentz
${\boldsymbol{\cal{L}}} ({\bf{r}})$ field.  The corresponding
components of the total energy $E$ as obtained from these fields are
also given in this section.  In Sect. V we demonstrate the
satisfaction of the `Quantal Newtonian'  first law by these fields.
The analytical and semi-analytical expressions of these properties
are given in Appendix A.  We employ this example of the triplet
$2^{3} S$ state in Sect VI to explain the self-consistent nature of
the Schr{\"o}dinger-Pauli equation.  Concluding remarks are made in
Sect VII with regard to the insights and numerous properties of this
triplet state as derived via the new perspective.


\section{Triplet $2^{3} S$ State Wave Function}

In the symmetric gauge ${\bf{A}} ({\bf{r}}) = \frac{1}{2}
{\boldsymbol{\cal{B}}} ({\bf{r}}) \times {\bf{r}}$, with the
magnetic field in the $z$-direction ${\boldsymbol{\cal{B}}}
({\bf{r}}) = B \hat{\bf{i}}_{z}$, the Schr{\"o}dinger-Pauli equation
Eq. (2) can be solved for the triplet $2^{3} S$ state of the $2$D
$2$-electron quantum dot in closed analytical form for a denumerably
infinite set of effective oscillator frequencies $\Omega^{2} =
\omega_{0}^{2} + \omega_{L}^{2}$ or effective force constant
$k_\mathrm{eff} = \Omega^{2}$, where $\omega_{L} = B/2c$ is the
Larmor frequency. Effective atomic units are employed:
$e^{2}/\epsilon = \hbar = m^{\star} = c = 1$.  The effective Bohr
radius is $a_{0}^{\star} = a_{0} (m/m^{\star})$, where $m$ is the
free electron mass. The effective energy unit is $(a.u.)^{\star} =
(a.u.)(m^{\star}/m \epsilon^{2})$. The wave function $\psi
({\bf{x}}_{1} {\bf{x}}_{2})$ for this excited state is a product of
a spatial $\Psi ({\bf{r}}_{1} {\bf{r}}_{2})$ and spin $\chi
(\sigma_{1} \sigma_{2})$ component:
\begin{equation}
\psi ({\bf{x}}_{1} {\bf{x}}_{2}) = \Psi ({\bf{r}}_{1} {\bf{r}}_{2})
\chi ({\bf{\sigma}}_{1} {\bf{\sigma}}_{2}).
\end{equation}
The spatial component $\Psi ({\bf{r}}_{1} {\bf{r}}_{2})$ is
\begin{eqnarray}
\Psi ({\bf{r}}_{1} {\bf{r}}_{2}) = N e^{i m \theta} e^{- \Omega
(r^{2}_{1} + r^{2}_{2})/2} \big[ | {\bf{r}}_{2} - {\bf{r}}_{1} | +
c_{2} | {\bf{r}}_{2} - {\bf{r}}_{1} |^{2} \nonumber \\
+ c_{3} | {\bf{r}}_{2} - {\bf{r}}_{1} |^{3}  + c_{4} | {\bf{r}}_{2}
- {\bf{r}}_{1} |^{4} \big],
\end{eqnarray}
where the normalization constant $N = 0.022466$; the angular quantum
number $m = 0, \pm 1, \pm 2, \ldots$ is chosen to be $m = +1$, the
coefficients $c_{2} = \frac{1}{3}$; $c_{3} = - 0.059108$; $c_{4} = -
0.015884$; the effective force constant $k_\mathrm{eff} = 0.072217$;
the angle $\theta$ is that of the relative coordinate vector
${\bf{u}} = {\bf{r}}_{2} - {\bf{r}}_{1}$; and ${\bf{r}}_{1} = (r_{1}
\theta_{1}), {\bf{r}}_{2} = (r_{2} \theta_{2})$.

The wave function $\psi ({\bf{x}}_{1} {\bf{x}}_{2})$ is of course
antisymmetric in an interchange of the coordinates ${\bf{x}}_{1}$
and ${\bf{x}}_{2}$.  Since the spin component $\chi
({\bf{\sigma}}_{1} {\bf{\sigma}}_{2})$ for the triplet state is
symmetric in an interchange of the coordinates ${\bf{\sigma}}_{1}$
and ${\bf{\sigma}}_{2}$, the spatial component $\Psi ({\bf{r}}_{1}
{\bf{r}}_{2})$ is antisymmetric in an interchange of ${\bf{r}}_{1}$
and ${\bf{r}}_{2}$, \emph{i.e}. $\Psi ({\bf{r}}_{1} {\bf{r}}_{2}) =
- \Psi ({\bf{r}}_{2} {\bf{r}}_{1})$.

The spatial part of the wave function $\Psi ({\bf{r}}_{1}
{\bf{r}}_{2})$ has many properties, and its structure is of interest
in its own right.  Here we exhibit some of these properties.  Other
properties are simply stated.  (They will be described in greater
detail elsewhere \cite{19} in a comparison with the wave function of
an excited singlet state of the quantum dot.)  The salient features
of $\Psi ({\bf{r}}_{1} {\bf{r}}_{2})$ are the following:

1.  In Figs. 1 - 4 we plot the function $\Psi ({\bf{r}}_{1}
{\bf{r}}_{2})$ as a function of $r_{1}$ and $r_{2}$ for different
$\theta_{1}$ and $\theta_{2}$. In each figure, panel (a) corresponds
to the real part of $\Psi ({\bf{r}}_{1} {\bf{r}}_{2})$, and panel
(b) to its imaginary part. Observe that in Fig. 1 for $\theta_{1} =
\theta_{2} = 0^{\circ}$, the function $\Psi ({\bf{r}}_{1}
{\bf{r}}_{2})$ is real.  As $\theta_{1}$ increases to $\theta_{1} =
45^{\circ}$ in Fig. 2, the real part shrinks and the imaginary part
becomes finite. For fixed $\theta_{1} = 45^{\circ}$ and increasing
$\theta_{2} = 60^{\circ}, 90^{\circ}$ as in Figs. 3 and 4,
respectively, the real part continues to diminish whilst the
imaginary part increases in magnitude.

2.  For an $N$ particle system, the coalescence condition \cite{8}
in $D$ dimensions of $2$ particles of masses $m_{1}$ and $m_{2}$,
and charges $Z_{1}$ and $Z_{2}$, (with the spin index suppressed) is
\begin{eqnarray}
\psi ({\bf{r}}_{1}, {\bf{r}}_{2}, \ldots, {\bf{r}}_{N}) = \psi
({\bf{r}}_{2}, {\bf{r}}_{2}, {\bf{r}}_{3}, \ldots, {\bf{r}}_{N})
\bigg( 1 + \frac{2 Z_{1} Z_{2} \mu_{1 2}} {D-1} u \bigg)
\nonumber\\
+ {\bf{u}} \cdot {\bf{C}} ({\bf{r}}_{2}, {\bf{r}}_{3}, \ldots,
{\bf{r}}_{N}),
\end{eqnarray}
where $\mu_{12} = m_{1} m_{2}/ m_{1} + m_{2}$ is the reduced mass,
and ${\bf{C}} ({\bf{r}}_{2}, {\bf{r}}_{3}, \ldots, {\bf{r}}_{N})$ is
an undetermined vector. This is the integral form of the \emph{cusp
coalescence} condition. It is equally valid when the wave function
vanishes at the point of coalescence, \emph{i.e}. when $\psi
({\bf{r}}_{2}, {\bf{r}}_{2}, {\bf{r}}_{3}, \ldots, {\bf{r}}_{N}) =
0$, and is then referred to as the \emph{node coalescence}
condition. The wave function $\psi ({\bf{x}}_{1} {\bf{x}}_{2})$ for
the triplet state via its spatial component $\Psi ({\bf{r}}_{1}
{\bf{r}}_{2})$ satisfies the node electron-electron coalescence
condition.

3.  The function $\Psi ({\bf{r}}_{1} {\bf{r}}_{2})$ exhibits the
following nodes.

(a) There is a node at the origin.  This is evident in Figs. 1 - 4
for both the cases of $\theta_{1} = \theta_{2}$ and $\theta_{1} \neq
\theta_{2}$. This is because the probability of $2$ electrons of the
same spin being at the same position in space at ${\bf{r}}_{1} =
{\bf{r}}_{2} = 0$ is zero as a result of the Pauli exclusion
principle. Observe also that the parity of the function $\Psi
({\bf{r}}_{1} {\bf{r}}_{2})$ about the origin is odd.

(b)  There is a node \cite{19} at all points of electron-electron
coalescence, again as a consequence of the Pauli exclusion
principle.  The function $\Psi ({\bf{r}}_{1} {\bf{r}}_{2})$ has odd
parity about \emph{all} these points of coalescence.

(c)  The real part of $\Psi ({\bf{r}}_{1} {\bf{r}}_{2})$ has a node
\cite{19} when the projections of the vectors ${\bf{r}}_{1}$ and
${\bf{r}}_{2}$ on the x-axis are the same.  The function $\Psi
({\bf{r}}_{1} {\bf{r}}_{2})$ is then purely imaginary.  The parity
of the $\Psi ({\bf{r}}_{1} {\bf{r}}_{2})$ is odd about the line
${\bf{r}}_{2} = (cos \theta_{1}/cos \theta_{2}) {\bf{r}}_{1}$.

(d)  The imaginary part of $\Psi ({\bf{r}}_{1} {\bf{r}}_{2})$ has a
node \cite{19} when the projections of the vectors ${\bf{r}}_{1}$
and ${\bf{r}}_{2}$ on the y-axis are the same. The wave function is
then real.  The parity of $\Psi ({\bf{r}}_{1} {\bf{r}}_{2})$ is odd
about the line ${\bf{r}}_{2} = (sin \theta_{1}/sin \theta_{2})
{\bf{r}}_{1}$ .

(e)  There is a node of $\Psi ({\bf{r}}_{1} {\bf{r}}_{2})$ as a
result of it being a first excited state.  These nodes are located
where $\Psi ({\bf{r}}_{1} {\bf{r}}_{2})$ is zero along the lines at
non-zero values of $r_{1}$ and $r_{2}$ as shown in Figs. 1-4. There
is no parity of $\Psi ({\bf{r}}_{1} {\bf{r}}_{2})$ about this node.


\section{QUANTAL SOURCES}

In this section we describe the various quantal sources for the
fields that satisfy the `Quantal Newtonian' first law.  As the spin
and spatial coordinates are separable, and the corresponding spin
$\chi ({\bf{\sigma}}_{1} {\bf{\sigma}}_{2})$ and spatial $\Psi
({\bf{r}}_{1} {\bf{r}}_{2})$ components are separately normalized,
the quantal sources are simply expectation values taken with respect
to the spatial component.  The analytical and semi-analytical
expressions for the sources are given in Appendix A.

\emph{\textbf{(i)  Electron Density} $\rho ({\bf{r}})$}

The electron density $\rho ({\bf{r}})$ is the expectation value
\begin{equation}
\rho ({\bf{r}}) = \langle \Psi | \hat{\rho}  ({\bf{r}}) | \Psi
\rangle,
\end{equation}
where the density operator $\hat{\rho}  ({\bf{r}})$ is
\begin{equation}
\hat{\rho}  ({\bf{r}}) = \sum_{k} \delta ({\bf{r}}_{k} - {\bf{r}}).
\end{equation}
In Fig. 5a the electron density $\rho ({\bf{r}})$ is plotted.  It is
spherically symmetric about the origin, and exhibits shell
structure.  As a consequence of the binding potential being
harmonic, the density is finite at the origin and does not exhibit a
cusp there.  It is a \emph{local} or \emph{static} property in that
its overall structure remains unchanged as the electron position is
varied.  In Fig. 5b, the radial probability density $r \rho
({\bf{r}})$ is plotted, and again shell structure is clearly evident
in the shoulder.

\emph{\textbf{(ii)  Physical Current Density} ${\bf{j}}({\bf{r}})$}

The physical current density ${\bf{j}}({\bf{r}})$, a \emph{local}
property, is the expectation value
\begin{equation}
{\bf{j}} ({\bf{r}}) = \langle \Psi | \hat{\bf{j}}  ({\bf{r}}) | \Psi
\rangle,
\end{equation}
where the current density operator  $\hat{\bf{j}}  ({\bf{r}})$ is
the sum of its paramagnetic $\hat{\bf{j}}_{p}  ({\bf{r}})$,
diamagnetic $\hat{\bf{j}}_{d}  ({\bf{r}})$, and magnetization
$\hat{\bf{j}}_{m}  ({\bf{r}})$ current density components:
\begin{equation}
\hat{\bf{j}} ({\bf{r}}) = \hat{\bf{j}}_{p}  ({\bf{r}}) +
\hat{\bf{j}}_{d}  ({\bf{r}}) + \hat{\bf{j}}_{m}  ({\bf{r}}),
\end{equation}
with
\begin{eqnarray}
\hat{\bf{j}}_{p}  ({\bf{r}}) &=&  \frac{1}{2} \sum_{k} \bigg[ \hat
{\bf{p}}_{k} \delta ({\bf{r}}_{k} - {\bf{r}}) + \delta ({\bf{r}}_{k}
- {\bf{r}}) \hat {\bf{p}}_{k} \bigg], \\
\hat{\bf{j}}_{d}  ({\bf{r}}) &=& \hat{\rho} ({\bf{r}}) {\bf{A}}
({\bf{r}}), \\
\hat{\bf{j}}_{m}  ({\bf{r}}) &=&  - {\boldsymbol{\nabla}} \times
\hat{\bf{m}} ({\bf{r}})
\end{eqnarray}
and the magnetization density $\hat{\bf{m}} ({\bf{r}})$ operator is
\begin{equation}
\hat{\bf{m}} ({\bf{r}}) = \sum_{k} {\bf{s}}_{k} \delta ({\bf{r}}_{k}
- {\bf{r}}).
\end{equation}
The paramagnetic current density ${\bf{j}}_{p}  ({\bf{r}})$ may also
be defined via the quantal source of the single-particle density
matrix $\gamma ({\bf{r r}}')$ as
\begin{equation}
{\bf{j}}_{p}  ({\bf{r}}) = \frac{1}{2} \bigg [{\boldsymbol{\nabla}}'
- {\boldsymbol{\nabla}}'' \bigg] \gamma ({\bf{r}}' {\bf{r}}'') \bigg
|_{{\bf{r}}' = {\bf{r}}'' = r},
\end{equation}
where $\gamma ({\bf{r r}}')$  is defined below in subsection
(\emph{\textbf{iv}}). (The expression for $\hat{\bf{j}}_{p}
({\bf{r}})$ for this triplet state given in Appendix A is derived
independently through the definitions of Eqs. (13) and (17).)

In Figs. 6 - 9 panels (a), the physical current density ${\bf{j}}
({\bf{r}})$, and its paramagnetic ${\bf{j}}_{p}  ({\bf{r}})$,
diamagnetic ${\bf{j}}_{d}  ({\bf{r}})$, and magnetization
${\bf{j}}_{m}  ({\bf{r}})$ components, respectively, are plotted.
The diamagnetic component ${\bf{j}}_{d}  ({\bf{r}})$ which is the
only component that depends explicitly on the magnetic field is
plotted for a value of the Larmor frequency of $\omega_{L} = 0.1$.
Hence, the plot of the total current density ${\bf{j}} ({\bf{r}})$
is for $\omega_{L} = 0.1$.  Each density component is a function
solely of the radial component $r$, but points in the
$\hat{\bf{i}}_{\theta}$ direction. Hence, the divergence of each
component vanishes, and therefore ${\boldsymbol{\nabla}} \cdot
{\bf{j}} ({\bf{r}}) = 0$.

Observe that shell structure is clearly evident in the plot of the
current density ${\bf{j}} ({\bf{r}})$ (see Fig. 6a).  This structure
is also evident in the individual components (see Figs. 7a - 9a),
although their individual structures are different.  For the choice
of $\omega_{L} = 0.1$, the magnitude of the paramagnetic
${\bf{j}}_{p}  ({\bf{r}})$, diamagnetic ${\bf{j}}_{d}  ({\bf{r}})$,
and magnetization ${\bf{j}}_{m}  ({\bf{r}})$ components is
essentially the same. (Depending on the value of $\omega_{L}$, the
diamagnetic component ${\bf{j}}_{d}  ({\bf{r}})$ and thus ${\bf{j}}
({\bf{r}})$ can be made larger or smaller.)

In Figs. 6 - 9, panels (b), the flow line contours of each current
density component are plotted.  These contour lines are closest in
the regions of greater density.  Observe the difference in the
contours for each density component.  The circulation direction of
the component ${\bf{j}}_{p}  ({\bf{r}})$ depends explicitly on the
choice of angular momentum quantum number $m$. This is also the case
for ${\bf{j}}_{m}  ({\bf{r}})$ whose dependency on $m$ is via the
electronic density $\rho_{\uparrow \uparrow}$ (corresponding to m =
1) or $\rho_{\downarrow \downarrow}$ (corresponding to $m = - 1$).
The circulation direction of these two current densities
${\bf{j}}_{p}  ({\bf{r}})$ and ${\bf{j}}_{m}  ({\bf{r}})$ is always
the same, but the direction depends upon whether $m = 1$ or $m =
-1$. On the other hand, the diamagnetic current density
${\bf{j}}_{d}  ({\bf{r}})$ does not depend on $m$. Thus, its
circulation can be either in the same or opposite direction to that
of ${\bf{j}}_{p}  ({\bf{r}})$ depending on the value of $m$. For our
choice of $m = 1$, the circulation direction for ${\bf{j}}_{p}
({\bf{r}})$, ${\bf{j}}_{d}  ({\bf{r}})$, and ${\bf{j}}_{m}
({\bf{r}})$ are all the same (counterclockwise). (The fact that the
circulation directions of ${\bf{j}}_{p}  ({\bf{r}})$ and
${\bf{j}}_{d}  ({\bf{r}})$ are the same, for the chosen value of
$m$, has been confirmed by an independent derivation related to the
contribution of the Lorentz ${\boldsymbol{\cal{L}}} ({\bf{r}})$ and
internal magnetic ${\boldsymbol{\cal{I}}}_{m} ({\bf{r}})$ fields to
the total energy.)

\emph{\textbf{(iii)  Pair-correlation Density} $g ({\bf{rr}}')$
\textbf{and the Fermi-Coulomb hole} $\rho_{xc} ({\bf{rr}}')$}

The pair-correlation density $g ({\bf{rr}}')$ is defined as the
ratio of the pair-correlation function $P ({\bf{rr}}')$ to the
density $\rho ({\bf{r}})$:
\begin{equation}
g ({\bf{rr}}') = P ({\bf{rr}}')/\rho ({\bf{r}}),
\end{equation}
where $P ({\bf{rr}}')$ is the expectation value
\begin{equation}
P ({\bf{rr}}') = \langle \Psi | \hat{P} ({\bf{rr}}') | \Psi \rangle,
\end{equation}
with the pair operator defined as
\begin{equation}
\hat{P} ({\bf{rr}}') = \sideset{}{'}\sum_{k,\ell} \delta
({\bf{r}}_{k} - {\bf{r}}) \delta ({\bf{r}}_{\ell} - {\bf{r}}').
\end{equation}
The pair-correlation density $g ({\bf{rr}}')$ may also be written in
terms of its local and nonlocal components as
\begin{equation}
g ({\bf{rr}}') = \rho ({\bf{r}}') + \rho_{xc} ({\bf{r r}}'),
\end{equation}
where $\rho_{xc} ({\bf{r r}}')$ is the Fermi-Coulomb hole charge.

The pair-correlation density $g ({\bf{rr}}')$ and Fermi-Coulomb hole
$\rho_{xc} ({\bf{r r}}')$ are \emph{nonlocal} quantal sources in
that their structure changes as a function of the electron position.
This is demonstrated in Fig. 10 where $g ({\bf{rr}}')$ is plotted
for the following different electron positions: (a) the center of
the quantum dot at $r = 0$; (b) at $r = 0.5~a.u.$; (c) at $r =
1.0~a.u.$;(d) at $r = 1.5~a.u.$ Observe that in each figure, the
pair-correlation density vanishes at the electron position. This is
a consequence of the node coalescence condition satisfied by the
wave function.  Also note that except for the electron position at
the center of the quantum dot, $g ({\bf{rr}}')$ is not spherically
symmetric about the electron position.  In Fig. 11, the $g
({\bf{rr}}')$ is plotted for asymptotic positions of the electron:
(a) at $r = 8.0~ a.u.$; (b) at $r = 12.0~ a.u.$ For these asymptotic
positions, observe that the figures are very similar. This is a
reflection of the fact that for such asymptotic positions of the
electron, the nonlocal charge is becoming essentially static. Since
the total charge of the pair-correlation density $g ({\bf{rr}}')$ is
$1$ (obtained from $N - 1$), the asymptotic structure of the
electron-interaction field ${\boldsymbol{\cal{E}}}_\mathrm{ee}
({\bf{r}})$ derived from it via Coulomb's law is analytically known
(see Sect. IV). Since for an electron at the center of the quantum
dot, the density $g ({\bf{rr}}')$ is spherically symmetric about
this position(Fig. 10a), the field vanishes there.

The nonlocal structure of the Fermi-Coulomb hole $\rho_{xc} ({\bf{r
r}}')$ (see Fig. 12) can be obtained from Eq. (21).  The hole
represents the reduction in density at ${\bf{r}}'$ for an electron
at ${\bf{r}}$ due to the Pauli exclusion principle and Coulomb
repulsion. Although this structure differs significantly from that
of $g ({\bf{rr}}')$, its properties are similar.  Thus, at the
electron position, the hole is finite and continuous and has the
lowest value.  (There is no cusp at this point as is the case for a
singlet excited state.) For an electron position at the center of
the quantum dot, the hole is spherically symmetric about it, and
thus the Pauli-Coulomb field ${\boldsymbol{\cal{E}}}_\mathrm{xc}
({\bf{r}})$ vanishes at the origin.  The hole is not spherically
symmetric about the other electron positions.  As the hole becomes
an essentially static charge for asymptotic positions of the
electron, and since the total hole charge is $-1$, the asymptotic
structure of ${\boldsymbol{\cal{E}}}_\mathrm{xc} ({\bf{r}})$ is also
analytically known (see Sect. IV.).

\emph{\textbf{(iv)   Single-Particle Density Matrix} $\gamma
({\bf{rr}}')$}

The single-particle density matrix $\gamma ({\bf{rr}}')$, a
\emph{nonlocal} quantal source, is defined as the expectation value
\begin{equation}
\gamma ({\bf{rr}}') = \langle \Psi | \hat{\gamma} ({\bf{rr}}') |
\Psi \rangle,
\end{equation}
where the complex single-particle density matrix operator \cite
{20,21} is
\begin{equation}
\hat{\gamma} ({\bf{rr}}')  = \hat{A} + i \hat{B},
\end{equation}
\begin{eqnarray}
\hat{A} = \frac{1}{2} \sum_{k} \big[ \delta ({\bf{r}}_{k} -
{\bf{r}}) T_{k} ({\bf{a}}) + \delta ({\bf{r}}_{k} - {\bf{r}}') T_{k}
(- {\bf{a}}) \big], \\
\hat{B} = \frac{i}{2} \sum_{k} \big[ \delta ({\bf{r}}_{k} -
{\bf{r}}) T_{k} ({\bf{a}}) - \delta ({\bf{r}}_{k} - {\bf{r}}') T_{k}
(- {\bf{a}}) \big],
\end{eqnarray}
with $T_{k} ({\bf{a}})$ a translation operator such that $T_{k}
({\bf{a}}) \psi (\ldots {\bf{r}}_{k}, \ldots) = \psi (\ldots
{\bf{r}}_{k} + {\bf{a}}, \ldots )$ and ${\bf{a}} = {\bf{r}}' -
{\bf{r}}$. The operators $\hat{A}$ and $\hat{B}$ are Hermitian.  The
single-particle density matrix is the quantal source for all kinetic
related properties \cite{22} such as the kinetic energy tensor, the
kinetic energy density, the kinetic field, the kinetic energy, and
as noted above, the paramagnetic current density.

In the panels of Fig. 13, the $\gamma ({\bf{rr}}')$ for the triplet
state is plotted as the positions ${\bf{r}}$ and ${\bf{r}}'$ change
for (a) $\theta = \theta' = 0^{\circ}$; (b) $\theta = 0^{\circ}$,
$\theta' = 45^{\circ}$; (c) $ \theta = 0^{\circ}$, $\theta' =
60^{\circ}$; (d) $ \theta = 0^{\circ}$, $\theta' = 90^{\circ}$. The
nonlocal nature of $\gamma ({\bf{rr}}')$ is clearly evident as is
shell structure. Observe the change in the shoulder of $\gamma
({\bf{rr}}')$ as $\theta'$ changes from $0^{\circ}$ to $90^{\circ}$.
Also note that the $\gamma ({\bf{rr}}')$ exhibits nodes as a
consequence of the node in the wave function for this excited state
(point \#3e of Sect. II). Although the wave function exhibits a node
at the origin (point \# 3a), the $\gamma ({\bf{rr}}')$ is finite
there.  For the cross sections for which ${\bf{r}} = {\bf{r}}'$, one
obtains the density of Fig.5 since $\gamma ({\bf{rr}}) = \rho
({\bf{r}})$.


\begin{table}
\caption{Properties of the Triplet $2^{3} S$ state of the quantum
dot in a magnetic field.  The values are in effective atomic units
$(a.u.)^{\star}$} \setlength\tabcolsep{5pt}
\begin{tabular}{|c|c|}
\hline\noalign{\smallskip}
 Property & Value \\
   \noalign {\smallskip}
  \hline
 $T$ & 0.615577 \\ \hline
 $E_{H}$ & 0.755497\\ \hline
 $E_{xc}$ & -0.501339 \\ \hline
 $E_{ee}$ & 0.254158 \\ \hline
 $E_{es} + E_{mag}$ & 0.742657\\ \hline
 $E$ &  1.612391 \\ \hline
 $IP = E^{N-1} - E^{N}$ &  -1.343659 \\ \hline
 $\langle {\bf{r}}^{2} \rangle $ &  20.567403\\ \hline
 $\langle r \rangle$ &  5.823553 \\ \hline
 $\langle  1/r \rangle$ & 1.041717 \\ \hline
 $\langle \delta ({\bf{r}}) \rangle$ &  0.0555377 \\ \hline
 \noalign{\smallskip} \hline \noalign{\smallskip}
\end{tabular}\\
\noindent
\end{table}


\section{`Forces', Fields, and Energies}

The `forces' and fields derived from the quantal sources are
described next.  The contributions of the individual fields to the
total energy $E$ are given in Table I.  Various analytical and
semi-analytical expressions for the fields and components of the
energy are given in Appendix A.

\emph{\textbf{(i)  Electron-interaction}}

The electron-interaction field ${\boldsymbol{\cal{E}}}_\mathrm{ee}
({\bf{r}})$ is obtained from its quantal source, the
pair-correlation density $g({\bf{r r}}')$, via Coulomb's law, and
may be written (see Eq. (21)) in terms of its Hartree
${\boldsymbol{\cal{E}}}_\mathrm{H} ({\bf{r}})$ and Pauli-Coulomb
${\boldsymbol{\cal{E}}}_\mathrm{xc} ({\bf{r}})$ components:
\begin{eqnarray}
{\boldsymbol{\cal{E}}}_\mathrm{ee} ({\bf{r}}) &=& \int
\frac{g({\bf{r r}}') ({\bf{r}} - {\bf{r}}')} {| {\bf{r}} - {\bf{r}}'
|^{3}} d
{\bf{r}}' \\
&=& {\boldsymbol{\cal{E}}}_\mathrm{H} ({\bf{r}}) +
{\boldsymbol{\cal{E}}}_\mathrm{xc} ({\bf{r}}),
\end{eqnarray}
where
\begin{equation}
{\boldsymbol{\cal{E}}}_\mathrm{H} ({\bf{r}}) = \int \frac{\rho
({\bf{r}}') ({\bf{r}} - {\bf{r}}')} {| {\bf{r}} - {\bf{r}}' |^{3}} d
{\bf{r}}' ~~ ; ~~ {\boldsymbol{\cal{E}}}_\mathrm{xc} ({\bf{r}}) =
\int \frac{\rho_\mathrm{xc} ({\bf{r r}}') ({\bf{r}} - {\bf{r}}')} {|
{\bf{r}} - {\bf{r}}' |^{3}} d {\bf{r}}'.
\end{equation}
The fields may also be expressed in terms of their corresponding
`forces' ${\bf{e}}_\mathrm{ee} ({\bf{r}})$, ${\bf{e}}_\mathrm{H}
({\bf{r}})$, and ${\bf{e}}_\mathrm{xc} ({\bf{r}})$:
\begin{equation}
{\boldsymbol{\cal{E}}}_\mathrm{ee} ({\bf{r}}) = {\bf{e}}_\mathrm{ee}
({\bf{r}})/\rho ({\bf{r}}) ~ ; ~ {\boldsymbol{\cal{E}}}_\mathrm{H}
({\bf{r}}) = {\bf{e}}_\mathrm{H} ({\bf{r}})/\rho ({\bf{r}}) ~ ; ~
{\boldsymbol{\cal{E}}}_\mathrm{xc} ({\bf{r}}) = {\bf{e}}_\mathrm{xc}
({\bf{r}})/\rho ({\bf{r}}).
\end{equation}

In Fig. 14(a) and (b) we plot the various `forces' and fields,
respectively.  Shell structure is evident in the plots of both the
`forces' and fields. (For the `force' ${\bf{e}}_\mathrm{ee}
({\bf{r}})$ and field ${\boldsymbol{\cal{E}}}_\mathrm{ee}
({\bf{r}})$, the second shell becomes evident on an expanded scale.)
As the quantal sources $g({\bf{r r}}')$, $\rho ({\bf{r}})$,
$\rho_\mathrm{xc} ({\bf{rr}}')$ are all cylindrically symmetric for
an electron position at the origin (see Figs. 5a, 10a, 12a), all the
corresponding fields vanish there. Since for asymptotic positions of
the electron in the classically forbidden region, the nonlocal
sources $g({\bf{r r}}')$ and  $\rho_\mathrm{xc} ({\bf{rr}}')$ become
essentially static charge distributions (see Fig. 11), and the
density $\rho ({\bf{r}})$ is a static charge, the asymptotic
structure of the fields as $r \rightarrow \infty$ is known exactly:
${\boldsymbol{\cal{E}}}_\mathrm{ee} ({\bf{r}}) \sim  1/r^{2}$,
${\boldsymbol{\cal{E}}}_\mathrm{H} ({\bf{r}}) \sim 2/r^{2}$,
${\boldsymbol{\cal{E}}}_\mathrm{xc} ({\bf{r}}) \sim -1/r^{2}$.  That
the decay of these fields is such is clearly evident in Fig. 14(b).
Asymptotically, the `forces'(see Fig. 14(a)) all vanish as their
decay is faster than that of the density.

The electron-interaction $E_\mathrm{ee}$, Hartree $E_\mathrm{H}$,
and Pauli-Coulomb $E_\mathrm{xc}$ energies are then obtained in
integral virial form from the respective fields as (see Table I)
\begin{eqnarray}
E_\mathrm{ee} &=& \int \rho ({\bf{r}}) {\bf{r}} \cdot
{\boldsymbol{\cal{E}}}_\mathrm{ee} ({\bf{r}}) d {\bf{r}}, \\
E_\mathrm{H} &=& \int \rho ({\bf{r}}) {\bf{r}} \cdot
{\boldsymbol{\cal{E}}}_\mathrm{H} ({\bf{r}}) d {\bf{r}}, \\
E_\mathrm{xc} &=& \int \rho ({\bf{r}}) {\bf{r}} \cdot
{\boldsymbol{\cal{E}}}_\mathrm{xc} ({\bf{r}}) d {\bf{r}}.
\end{eqnarray}

\emph{\textbf{(ii) Kinetic}}

The quantal source for the kinetic `force' ${\bf{z}} ({\bf{r}})$,
field ${\boldsymbol{\cal{Z}}} ({\bf{r}})$, and energy $T$ is the
single-particle density matrix $\gamma ({\bf{rr}}')$. The field is
defined in terms of the `force' as
\begin{equation}
{\boldsymbol{\cal{Z}}} ({\bf{r}}) = {\bf{z}} ({\bf{r}})/\rho
({\bf{r}}),
\end{equation}
where in Cartesian coordinates
\begin{equation}
z_{\alpha} ({\bf{r}}) = 2 \sum_{\beta} \nabla_{\beta} t_{\alpha
\beta} ({\bf{r}} ; \gamma),
\end{equation}
and where the second-rank kinetic energy tensor
\begin{equation}
t_{\alpha \beta} ({\bf{r}} ; \gamma) = \frac{1}{4} \bigg[
\frac{\partial^{2}} {\partial r'_{\alpha} \partial r''_{\beta}} +
\frac{\partial^{2}} {\partial r'_{\beta} \partial r''_{\alpha}}
\bigg] \gamma ({\bf{r}}' {\bf{r}}'') \bigg|_{{\bf{r}}' = {\bf{r}}''
= r} .
\end{equation}
The kinetic energy in terms of the field ${\boldsymbol{\cal{Z}}}
({\bf{r}}) $ is
\begin{equation}
T = - \frac{1}{2} \int \rho ({\bf{r}})  {\bf{r}} \cdot
{\boldsymbol{\cal{Z}}} ({\bf{r}}) d {\bf{r}} .
\end{equation}
The kinetic `force' $z ({\bf{r}})$ and field ${\boldsymbol{\cal{Z}}}
({\bf{r}})$ are plotted in Fig. 15 (a) and (b), respectively.  Once
again, shell structure is evident. Whilst the `force' $z ({\bf{r}})$
decays and vanishes asymptotically, the field
${\boldsymbol{\cal{Z}}} ({\bf{r}})$ is singular in this region. Both
vanish at the origin. See Table I for the value of $T$.  (For the
derivation of the tensor $t_{\alpha \beta} ({\bf{r}} ; \gamma)$ and
the kinetic `force' ${\bf{z}} ({\bf{r}})$, see Appendix B.)

\emph{\textbf{(iii) Differential Density}}

The quantal source for the differential density `force' ${\bf{d}}
({\bf{r}})$ and field ${\boldsymbol{\cal{D}}} ({\bf{r}})$ is the
density $\rho ({\bf{r}})$.  The field ${\boldsymbol{\cal{D}}}
({\bf{r}})$ is defined as
\begin{equation}
{\boldsymbol{\cal{D}}} ({\bf{r}}) = {\bf{d}} ({\bf{r}})/\rho
({\bf{r}}),
\end{equation}
where
\begin{equation}
{\bf{d}} ({\bf{r}}) = - \frac{1}{4} {\boldsymbol{\nabla}} \nabla^{2}
\rho ({\bf{r}}).
\end{equation}
The `force' ${\bf{d}} ({\bf{r}})$ and field ${\boldsymbol{\cal{D}}}
({\bf{r}})$ are plotted in Fig. 16 (a) and (b), respectively.  Their
structure is similar to the kinetic case.  The `force' ${\bf{d}}
({\bf{r}})$ and field ${\boldsymbol{\cal{D}}} ({\bf{r}})$ exhibit
shell structure, they both vanish at the origin, the `force' decays
asymptotically, whereas the field is singular in that region.  There
is no direct contribution of this field to the energy, however, its
quantal source $\rho ({\bf{r}})$ is the source for the Hartree field
${\boldsymbol{\cal{E}}}_\mathrm{H} ({\bf{r}})$, and contributes to
the energy through every contribution of the other energy components
such as ${\boldsymbol{\cal{E}}}_\mathrm{ee} ({\bf{r}})$, $T$, etc.
(see Eqs. (30), (36), (47), and (48)).

\emph{\textbf{(iv) Lorentz, Internal Magnetic, and External
Electrostatic}}

The quantal source for the Lorentz and internal magnetic `forces'
$({\boldsymbol{\ell}} ({\bf{r}}), {\bf{i}}_{m} ({\bf{r}}))$ and
fields $({\boldsymbol{\cal{L}}} ({\bf{r}}),
{\boldsymbol{\cal{I}}}_{m} ({\bf{r}}))$ is the physical current
density ${\bf{j}} ({\bf{r}})$. The Lorentz field
${\boldsymbol{\cal{L}}} ({\bf{r}})$ is defined as
\begin{equation}
{\boldsymbol{\cal{L}}} ({\bf{r}}) = {\boldsymbol{\ell}}
({\bf{r}})/\rho ({\bf{r}}),
\end{equation}
where
\begin{equation}
{\boldsymbol{\ell}} ({\bf{r}}) = {\bf{j}} ({\bf{r}}) \times
{\boldsymbol{\cal{B}}} ({\bf{r}}),
\end{equation}
or in Cartesian coordinates
\begin{equation}
\ell_{\alpha} ({\bf{r}}) = \sum_{\beta} \big[ j_{\beta} ({\bf{r}})
\nabla_{\alpha} A_{\beta} ({\bf{r}}) - j_{\beta} ({\bf{r}})
\nabla_{\beta} A_{\alpha} ({\bf{r}}) \big ].
\end{equation}
The internal magnetic field ${\boldsymbol{\cal{I}}}_{m} ({\bf{r}})$
is defined as
\begin{equation}
{\boldsymbol{\cal{I}}}_{m} ({\bf{r}}) = {\bf{i}}_{m} ({\bf{r}})/\rho
({\bf{r}}),
\end{equation}
where in Cartesian coordinates
\begin{equation}
i_{m, \alpha} ({\bf{r}}) = \sum_{\beta} \nabla_{\beta} I_{\alpha
\beta} ({\bf{r}}),
\end{equation}
and where the second-rank tensor
\begin{equation}
I_{\alpha \beta} ({\bf{r}}) = \big[j_{\alpha} ({\bf{r}}) A_{\beta}
({\bf{r}}) + j_{\beta} ({\bf{r}}) A_{\alpha} ({\bf{r}}) \big] - \rho
({\bf{r}}) A_{\alpha} ({\bf{r}}) A_{\beta} ({\bf{r}}).
\end{equation}
We next define the field ${\boldsymbol{\cal{M}}}  ({\bf{r}})$ as
\begin{equation}
{\boldsymbol{\cal{M}}}  ({\bf{r}}) = - \big[ {\boldsymbol{\cal{L}}}
({\bf{r}}) + {\boldsymbol{\cal{I}}}_{m}  ({\bf{r}}) \big].
\end{equation}
Then, if ${\boldsymbol{\nabla}} \times {\boldsymbol{\cal{M}}}
({\bf{r}}) = 0$, as is the case in the present application, one can
define a \emph{path-independent} scalar magnetic potential $v_{m}
({\bf{r}})$ such that
\begin{equation}
{\boldsymbol{\cal{M}}}  ({\bf{r}}) = - {\boldsymbol{\nabla}} v_{m}
({\bf{r}})/e.
\end{equation}
Hence, the contribution to the energy $E_\mathrm{mag}$ of the sum of
the Lorentz and internal magnetic fields ${\boldsymbol{\cal{M}}}
({\bf{r}})$ is
\begin{equation}
E_\mathrm{mag} = \int \rho ({\bf{r}}) v_{m} ({\bf{r}}) d {\bf{r}}.
\end{equation}
In a similar manner, as the external electrostatic field
${\boldsymbol{\cal{E}}} ({\bf{r}}) = - {\boldsymbol{\nabla}} v
({\bf{r}})/e$ is curl free, the contribution to the energy
$E_\mathrm{es}$ due to this field is
\begin{equation}
E_\mathrm{es} = \int \rho ({\bf{r}}) v ({\bf{r}}) d {\bf{r}}.
\end{equation}
(The expression for $E_\mathrm{es}$ can also be obtained directly
from the Hamiltonian of Eq. (2) as the expectation value of the
operator $\hat{V}$.)

Both the Lorentz and internal magnetic `forces'
$({\boldsymbol{\ell}} ({\bf{r}}), {\bf{i}}_{m} ({\bf{r}}))$ and
fields (${\boldsymbol{\cal{L}}} ({\bf{r}}),
{\boldsymbol{\cal{I}}}_{m} ({\bf{r}}))$ depend on the strength of
the magnetic field. In Fig. 17, these `forces' and fields are
plotted for a value of the Larmor frequency of $\omega_{L} = 0.1$.
Again, observe that these properties exhibit shell structure.  The
`forces' Fig. 17 (a) vanish at the origin and asymptotically in the
classically forbidden region.  The fields Fig. 17 (b) vanish at the
origin, but are singular asymptotically.  For this triplet state of
the quantum dot, it turns out that
\begin{equation}
{\boldsymbol{\cal{M}}} ({\bf{r}}) = - \omega^{2}_{L} r
\hat{\bf{i}}_{r},
\end{equation}
and this linear function is also plotted in Fig. 17 (b).  It follows
from Eqs. (46) and (49) that (in $a. u.)^{\star}$
\begin{equation}
v_{m} ({\bf{r}}) = \frac{1}{2} \omega^{2}_{L} r^{2}.
\end{equation}
Thus, the sum of the electrostatic $E_\mathrm{es}$ and magnetostatic
$E_\mathrm{mag}$ energies is
\begin{eqnarray}
E_\mathrm{es} + E_\mathrm{mag} &=& \int \rho (r) \bigg\{ \frac{1}{2}
\bigg[ \omega^{2}_{0} + \omega^{2}_{L} \bigg] \bigg\} d {\bf{r}} \\
&=& \int \rho (r) \bigg[ \frac{1}{2} k_\mathrm{eff} r^{2} \bigg] d
{\bf{r}},
\end{eqnarray}
where $k_\mathrm{eff} = \omega_{0}^{2} +  \omega_{L}^{2} = 0.072217$
(see Sect. II). The value of this sum of energies is given in Table
I.

The total energy of this triplet $2^{3} S$ state can then be written
as
\begin{equation}
E = T + E_{H} + E_\mathrm{xc} + E_\mathrm{es} + E_\mathrm{mag} =
1.612391 ~ (a.u.)^{\star}.
\end{equation}
(This value is consistent with the eigenvalue for the triplet state
obtained by solution of the Schr{\"o}dinger-Pauli equation Eq. (1).)
The ionization potential defined as $IP = E^{N=1} - E^{N=2}$, where
$E^{N=1} = \Omega (n + 1); ~ n=0$, is also quoted in Table I. Note
that the same effective frequency $\Omega$ is employed in both terms
to determine the $IP$.  In addition to the values of these energy
components and the ionization potential, the values of the
expectations of the operators $\hat{O} = r^{2}, r, 1/r$, and
$\delta({\bf{r}})$ are also quoted. These latter expectation values
are related to various properties of the system such as the
diamagnetic susceptibility, the size of the `artificial atom', and
the electron density at the origin.


\section{`Quantal Newtonian' First Law}

For the triplet state of the quantum dot, the `Quantal Newtonian'
first law of Eq. (3) may be written in terms of the individual
fields, binding $\omega_{0}$ and Larmor $\omega_{L}$ frequencies ,
and the effective force constant $k_\mathrm{eff}$ as
\begin{eqnarray}
- k_\mathrm{eff} r &=& - [\omega^{2}_{0} + \omega^{2}_{L}] r \\
&=& - \omega^{2}_{0} r - [{\boldsymbol{\cal{L}}} ({\bf{r}}) +
{\boldsymbol{\cal{I}}}_{m} ({\bf{r}}) ] = -
{\boldsymbol{\cal{E}}}_{ee} ({\bf{r}}) + {\boldsymbol{\cal{Z}}}
({\bf{r}}) + {\boldsymbol{\cal{D}}} ({\bf{r}}).
\end{eqnarray}
These fields are plotted in Fig. 18.  In the figure, the Lorentz
${\boldsymbol{\cal{L}}} ({\bf{r}})$ and internal magnetic
${\boldsymbol{\cal{I}}}_{m} ({\bf{r}})$ fields, which are the only
two fields that depend on the magnetic field, are drawn for
$\omega_{L} = 0.1$. As shown in Fig. 17, the singularities in these
two fields cancel to lead to the linear function $\omega_{L}^{2} r$.
The singularities in the differential density
${\boldsymbol{\cal{D}}} ({\bf{r}})$ and kinetic
${\boldsymbol{\cal{Z}}} ({\bf{r}})$ fields also cancel to lead to a
linear function (see plot of ${\boldsymbol{\cal{D}}} ({\bf{r}}) +
{\boldsymbol{\cal{Z}}} ({\bf{r}})$ in Fig. 18).  On the addition of
the electron-interaction field $- {\boldsymbol{\cal{E}}}_{ee}
({\bf{r}})$ to ${\boldsymbol{\cal{D}}} ({\bf{r}}) +
{\boldsymbol{\cal{Z}}} ({\bf{r}})$, one obtains the linear function
$- k_\mathrm{eff} r$.  This then demonstrates the satisfaction of
the `Quantal Newtonian' first law by the various fields experienced
by each electron.


\section{Self-Consistent Nature of the Schr{\"o}dinger-Pauli Equation}

The example of the triplet state of the quantum dot in a magnetic
field can be employed to demonstrate the intrinsic self-consistent
nature of the Schr{\"o}dinger-Pauli equation.  Consider the
Schr{\"o}dinger-Pauli equation for the quantum dot written in its
generalized form (See Part I) (in effective atomic units)
\begin{equation}
\hat{H} [\Psi] \Psi = E [\Psi] \Psi,
\end{equation}
where the Hamiltonian $\hat{H} [\Psi]$ with \emph{unknown} binding
potential $v [\Psi] ({\bf{r}})$ is
\begin{equation}
\hat{H} [\Psi] = \frac{1}{2} \sum^{2}_{k=1} \big({\bf{p}}_{k} +
{\bf{A}} ({\bf{r}}_{k}) \big)^{2} + \sum^{2}_{k=1} {\bf{B}}
({\bf{r}}_{k}) \cdot {\bf{s}}_{k} + \frac{1}{2}
\sideset{}{'}\sum_{k,\ell=1}^{2} \frac{1}{|{\bf{r}}_{k} -
{\bf{r}}_{\ell} |} + \sum^{2}_{k=1} v [\Psi] ({\bf{r}}_{k}),
\end{equation}
with
\begin{equation}
v [\Psi] ({\bf{r}}) = \int^{\bf{r}}_{\infty} \big[
{\boldsymbol{\cal{E}}}_\mathrm{ee} ({\bf{r}}') -
{\boldsymbol{\cal{D}}} ({\bf{r}}') - {\boldsymbol{\cal{Z}}}
({\bf{r}}') - {\boldsymbol{\cal{L}}} ({\bf{r}}') -
{\boldsymbol{\cal{I}}}_{m}  ({\bf{r}}') \big] \cdot d
{\boldsymbol{\ell}}'.
\end{equation}
(The above equations are written in terms of the spatial part $\Psi
({\bf{r}}_{1} {\bf{r}}_{2})$ of the wave function.)  In the
symmetric gauge ${\bf{A}} ({\bf{r}}) = \frac{1}{2}
{\boldsymbol{\cal{B}}} ({\bf{r}}) \times {\bf{r}}$ ;
${\boldsymbol{\cal{B}}} ({\bf{r}}) =  {\cal{B}} {\bf{i}}_{z}$, with
the assumption of cylindrical symmetry, let us assume the form of a
trial input wave function to be
\begin{equation}
\Psi ({\bf{r}}_{1} {\bf{r}}_{2}) = Ne^{i m \theta} e^{- \Omega(R^{2}
+ \frac{1}{4} u^{2})} \big[u + c_{2} u^{2} + c_{3} u^{3} + c_{4}
u^{4} \big]
\end{equation}
where ${\bf{r}} = ({\bf{r}}_{1} + {\bf{r}}_{2})/2 ~;~ {\bf{u}} =
|{\bf{r}}_{2} - {\bf{r}}_{1} |$; the angular momentum quantum number
$m=1$; the Larmor frequency $\omega_{L} = 0.1$; and $N, \Omega
\equiv \sqrt{k_\mathrm{eff}}, c_{2}, c_{3}, c_{4}$ are
\emph{unknown} constants.  As a consequence of cylindrical symmetry,
let us assume all the individual fields are conservative.

For an assumed choice of the values of the constants, employ the
input $\Psi ({\bf{r}}_{1} {\bf{r}}_{2})$ to determine the various
fields, and from them the potential $v [ \Psi ] ({\bf{r}})$.
Substitute this $v [ \Psi ] ({\bf{r}})$ into Eq. (56) to solve for
$\Psi ({\bf{r}}_{1} {\bf{r}}_{2})$ with new coefficients, and repeat
the iterative procedure.  At each iteration one also obtains the
corresponding $E [\Psi ]$.  (We reiterate that what is meant by the
functional $v [ \Psi ] ({\bf{r}})$ is that for each different
$\Psi$, one obtains a different function $v [ \Psi ] ({\bf{r}})$.)

Suppose at the end of a particular iteration, the values of the
coefficients turn out to be $N = 0.022466$, $c_{2} = 0.33333$,
$c_{3} = - 0.059108$, $c_{4} = - 0.015884$; $\Omega^{2} \equiv
k_\mathrm{eff} = 0.072217$.  On substituting the $v ({\bf{r}})$ of
Eq. (58) into the Schr{\"o}dinger-Pauli equation Eq. (56) and
solving, one obtains a wave function $\Psi ({\bf{r}}_{1}
{\bf{r}}_{2})$ with the same coefficients.  Thus, this constitutes
the final iteration of the self-consistent procedure, thereby
leading to the exact wave function $\Psi ({\bf{r}}_{1}
{\bf{r}}_{2})$ and energy $E$.  Hence, via the self-consistency
procedure one obtains the Hamiltonian, the wave function, and eigen
energy.

An examination of the final result of the sum of $-
{\boldsymbol{\cal{E}}}_\mathrm{ee} ({\bf{r}})$,
${\boldsymbol{\cal{D}}} ({\bf{r}})$, and ${\boldsymbol{\cal{Z}}}
({\bf{r}})$ fields will show it to be a linear function of slope $-
k_\mathrm{eff} = - 0.072217$ (see Fig. 18), as follows
\begin{equation}
- {\boldsymbol{\cal{E}}}_\mathrm{ee} ({\bf{r}}) +
{\boldsymbol{\cal{D}}} ({\bf{r}}) + {\boldsymbol{\cal{Z}}}
({\bf{r}}) = - k_\mathrm{eff} r.
\end{equation}
As the individual fields are conservative, the sum of the magnetic
fields ${\boldsymbol{\cal{M}}} ({\bf{r}}) = -
[{\boldsymbol{\cal{L}}} ({\bf{r}}) + {\boldsymbol{\cal{I}}}_{m}
({\bf{r}}) ]$ is such that ${\boldsymbol{\nabla}} \times
{\boldsymbol{\cal{M}}} ({\bf{r}}) = 0$.  Then these fields can be
associated with a magnetic scalar potential $v_{m} ({\bf{r}})$
through
\begin{equation}
v_{m} [\Psi] ({\bf{r}}) = - \int^{\bf{r}}_{\infty}
{\boldsymbol{\cal{M}}} ({\bf{r}}') \cdot d {\boldsymbol{\ell}}'.
\end{equation}
As such, Eq. (58) can be rearranged to read
\begin{eqnarray}
v [\Psi] ({\bf{r}}) + v_{m} [\Psi] ({\bf{r}}) &\equiv&
v_\mathrm{eff} [\Psi] ({\bf{r}}) \nonumber \\
&=& \int^{\bf{r}}_{\infty} \big[ {\boldsymbol{\cal{E}}}_\mathrm{ee}
({\bf{r}}') - {\boldsymbol{\cal{D}}} ({\bf{r}}') -
{\boldsymbol{\cal{Z}}} ({\bf{r}}') \big] \cdot d
{\boldsymbol{\ell}}'.
\end{eqnarray}
On substituting Eq. (60) in Eq. (62) the effective potential
$v_\mathrm{eff} ({\bf{r}})$ turns out to be harmonic as
\begin{equation}
v_\mathrm{eff} ({\bf{r}}) = \frac{1}{2} k_\mathrm{eff} r^{2},
\end{equation}
with $k_\mathrm{eff}$ as given above.

A further examination of the final result of the field
${\boldsymbol{\cal{M}}} ({\bf{r}})$ shows that the corresponding
magnetic scalar potential  $v_{m} ({\bf{r}})$ is also harmonic with
Larmor frequency $\omega_{L} = 0.1$ as
\begin{equation}
v_{m} ({\bf{r}}) = \frac{1}{2} \omega^{2}_{L} r^{2}.
\end{equation}

Consequently, since both $v_\mathrm{eff} ({\bf{r}})$ and $v_{m}
({\bf{r}})$ are harmonic, their difference which is the binding
external electrostatic potential $v ({\bf{r}})$, must also be
harmonic, with some angular frequency $\omega_{0}$ as
\begin{equation}
v ({\bf{r}}) = \frac{1}{2} \omega^{2}_{0} r^{2},
\end{equation}
where $\omega^{2}_{0} = k_\mathrm{eff} - \omega^{2}_{L} = 0.062217$.
Thus, via the self-consistent procedure, the unknown external
potential $v ({\bf{r}})$ is determined to be harmonic.  (Note that
for a different value of the magnetic field ${\boldsymbol{\cal{B}}}
({\bf{r}})$ or equivalently Larmor frequency $\omega_{L}$, one would
also obtain a $v ({\bf{r}})$ that is harmonic, but with a different
angular frequency $\omega_{0}$.  However, the value of the effective
force constant $k_\mathrm{eff}$ will remain unchanged.)


\section{Concluding Remarks}

The purpose of this paper is to demonstrate by example the new
perspective of Schr{\"o}dinger-Pauli theory of electrons explained
in Part I.  The perspective complements the traditional description
of quantum mechanics in that it leads to a deeper understanding of
the physical system.  The perspective is that of the
\emph{individual} electron via its equation of motion, or
equivalently, in the stationary state case, the `Quantal Newtonian'
first law.  The law is in terms of `classical' fields whose sources
are quantum-mechanical expectation values of Hermitian operators
taken with respect to the wave function.  The structure of the
quantal sources is predictive of the structure of the fields.  One
new insight obtained via the law is that in addition to the external
fields, each electron also experiences an internal field.  And this
internal field is comprised of components that are representative of
intrinsic properties of the system -- the correlations due to the
Pauli Exclusion Principle and Coulomb repulsion, the electronic
density, kinetic effects, and an internal magnetic field component
dependent on the electronic current density.  Thus, in the presence
of a magnetic field, not only does each electron experience a
Lorentz field, but also an internal magnetic field.  It further
experiences a kinetic field. The total energy and its components can
also be described in terms of these fields.  The field description
makes for an understanding of the quantum system that is tangible.
As explained in Part I, the fields are deterministic.  The sources
of these fields are probabilistic in that they are
quantum-mechanical expectation values.  The quantal sources are both
\emph{local} and \emph{nonlocal}, and that is why one must first
determine fields from which then potentials and energies can be
obtained. These new understandings then enhance the traditional
quantum-mechanical description of a physical system.

A second new understanding is that all the quantal sources and
fields are related to each other in a self-consistent manner.  This
is because as shown in Part I, the Schr{\"o}dinger-Pauli equation
can be written in a generalized form in terms of the various fields
descriptive of the system.  Written in this manner, it  shows the
Schr{\"o}dinger-Pauli equation to be an intrinsically
self-consistent one.  This then provides a self-consistent procedure
for the solution of the Schr{\"o}dinger-Pauli equation.

The quantal-source -- field perspective of the Schr{\"o}dinger-Pauli
theory via the `Quantal Newtonian' first law is applied in this work
to the triplet $2^{3} S$ state of a $2$-dimensional $2$-electron
quantum dot in a magnetic field.  The purpose of the application is
to demonstrate each facet of the new formalism.  Hence, we first
study the \emph{local} quantal sources of the density $\rho
({\bf{r}})$, and the current density ${\bf{j}} ({\bf{r}})$ and its
paramagnetic ${\bf{j}}_{p} ({\bf{r}})$, diamagnetic ${\bf{j}}_{d}
({\bf{r}})$, and magnetization ${\bf{j}}_{m} ({\bf{r}})$ components,
and their circulation; and then the \emph{nonlocal} sources of the
pair-correlation density $g ({\bf{rr}}')$, the Fermi-Coulomb hole
charge distribution $\rho_\mathrm{xc} ({\bf{rr}}')$, and the
single-particle density matrix $\gamma ({\bf{rr}}')$.  These sources
give rise to fields experienced by each electron: the
electron-interaction field ${\boldsymbol{\cal{E}}}_\mathrm{ee}
({\bf{r}})$ and its Hartree ${\boldsymbol{\cal{E}}}_{H} ({\bf{r}})$
and Pauli-Coulomb ${\boldsymbol{\cal{E}}}_\mathrm{xc} ({\bf{r}})$
components; the kinetic field ${\boldsymbol{\cal{Z}}} ({\bf{r}})$;
the differential density field ${\boldsymbol{\cal{D}}} ({\bf{r}})$;
the Lorentz ${\boldsymbol{\cal{L}}} ({\bf{r}})$ and internal
magnetic ${\boldsymbol{\cal{I}}}_{m} ({\bf{r}})$ fields.  These
fields exhibit characteristics of the system such as shell
structure.  They satisfy the `Quantal Newtonian' first law. Together
with the external electrostatic field ${\boldsymbol{\cal{E}}}
({\bf{r}})$, these fields then give rise to the components of the
total energy: the electron-interaction $E_\mathrm{ee}$, Hartree
$E_{H}$, Pauli-Coulomb $E_\mathrm{xc}$, kinetic $T$, the external
electrostatic $E_\mathrm{es}$, and the magnetostatic
$E_\mathrm{mag}$ with $E_\mathrm{mag}$ the sum of the Lorentz and
internal magnetic energy.

Finally, the example allows for the demonstration of the
self-consistency procedure of the Schr{\"o}dinger-Pauli equation.
Thus, it is shown how in the last iteration of the self-consistent
procedure, the exact wave function, eigen energy, and the external
binding potential and therefore the Hamiltonian are determined.

\appendix

\section{Expressions for Properties of the Triplet $2^3S$ State of a
Two-Electron Quantum Dot in a Magnetic Field}

In this Appendix we provide the closed-form analytical and
semi-analytical expressions for various properties of the $2^{3} S$
state of the quantum dot in a magnetic field.  In these expressions
the zeroth\!- and first-order modified Bessel functions $I_{0} (x)$
and $I_{1} (x)$ appear.  The general Bessel function $I_{\nu} (x)$
is defined as \cite{23}
\begin{equation}
I_{\nu} (x) = \sum_{n=0}^{\infty} \frac{1} {n! \Gamma (n + \nu + 1)}
\bigg(\frac{1}{2} x \bigg)^{2n + \nu},
\end{equation}
where the Gamma function $\Gamma (n)$ is \cite{23}
\begin{equation}
\Gamma (n) = \int^{\infty}_{0} t^{n-1} e^{-t} d t ~~; ~~ n>0.
\end{equation}
Also the parameter $\Omega$ is related to the binding frequency
$\omega_{0}$, the Larmor frequency $\omega_{L}$, and the effective
force constant $k_\mathrm{eff}$ as
\begin{equation}
\Omega = \sqrt{k_\mathrm{eff}} = \sqrt{\omega^{2}_{0} +
\omega^{2}_{L}} = 0.268732
\end{equation}
Other constants that appear in the expressions are
\begin{eqnarray}
A &=& \frac{1}{3} ~ ; ~ B = \frac{1}{8} (\frac{1}{3} - 3\Omega) = -
0.059108 \nonumber \\
C &=& \frac{1}{360} (1- 25 \Omega) = - 0.015884.
\end{eqnarray}

\textbf{Electron density $\rho({\bf r})$}
\begin{eqnarray}
&\rho(\emph r) = 4 \pi N^{2} e^{- 2 \Omega r^{2}} \int^{\infty}_{0}
e^{- \Omega x^{2}} x \big[ x + c_{2} x^{2} + c_{3} x^{3} + c_{4}
x^{4} \big]^{2} I_{0} (2 \Omega r x ) d x
\nonumber \\
&= \frac{N^2}{4\Omega^9} \bigg\{8 \pi \Omega^4 e^{-\Omega r^2}\big(
K_1 + L_1  r^2+M_1 r^4+N_1 r^6+O_1 r^8 \big) +
\nonumber \\
&\pi^{3/2}  \Omega^{9/2} \ e^{- \frac{3}{2}\Omega r^2} \bigg [K_2
+L_2  r^2+M_2r^4+N_2 r^6 +O_2 r^8\big) \  I_o\big( \Omega r^2 /2
\big) +
\nonumber \\
&2 \Omega \big( L_3  r^2+M_3 r^4+N_3 r^6+O_3 r^8 \big) \  I_1\big(
\Omega r^2 /2 \big)  \bigg]  \bigg\},&
\end{eqnarray}
where
\begin{eqnarray}
\ N&=&0.02246632108,
\nonumber\\
\ K_1 &=& 24 C^2 +(6 B^2 + 12 A C)\  \Omega + 2 A^2\  \Omega^2,
\nonumber \\
L_1 &=& 96\  C^2\  \Omega +  (18 B^2 + 36 A C) \ \Omega^2+(4 A^2 + 8
B)\ \Omega^3 + \Omega^4,
\nonumber \\
M_1 &=& 2\  C^2\  \Omega^2 + (9 B^2 + 18 A C) \ \Omega^3 + (A^2 + 2
B) \ \Omega^4,
\nonumber \\
N_1 &=& 16 \ C^2\  \Omega^3 + (B^2 + 2 A C) \ \Omega^4,
\nonumber \\
O_1&=& C^2  \ \Omega^4,
\nonumber\\
K_2 &=& 105 B C + 30 (A B + C)\  \Omega + 12 A \  \Omega^2,
\nonumber\\
L_2 &=& 420 B C \ \Omega + 90 (A B + C)\  \Omega^2 + 24  A\
\Omega^3,
\nonumber\\
M_2 &=& 376 B C \ \Omega^2 + 56 (A B + C) \  \Omega^3 + 8 A \
\Omega^4,
\nonumber\\
N_2 &=& 104 B C \  \Omega^3 + 8 (A B + C) \ \Omega^4,
\nonumber \\
O_2&=& 8 B C  \ \Omega^4,
\nonumber\\
L_3 &=& 88 B C + 23 (A B +C) \ \Omega + 8 A \  \Omega^2 ,
\nonumber\\
M_3 &=& 142 B C \ \Omega + 24 (A B + C) \ \ \Omega^2 + 4 A \
\Omega^3,
\nonumber\\
N_3 &=& 48 B C \ \Omega^2 + 4 (A B +C)  \  \Omega^3,
\nonumber\\
O_3&=& 4 B C  \ \Omega^3.
\end{eqnarray}
The asymptotic structure of $\rho(\bf{r})$  near the center of the
quantum dot, and in the classically forbidden region, respectively,
is as follows:
\begin{equation}
\rho(r) \ \substack{ \  \\  \\  \sim \\{r \to 0}} \ 0.0555 -
0.00625\ r^2 - 0.000230\ r^4 + \ldots
\end{equation}
with $\rho(0) = 0.0555377\  a.u.$,
\begin{eqnarray}
\rho(r) \ \substack{ \  \\  \\  \sim \\{r \to \infty}} &e^{-\Omega
r^2} \ 10^{-5} \ (8.03 \ r^4 + 12.6\ r^5+ 9.35 \ r^6+2.22 \ r^7+
0.298 \ r^8 + \ldots )
\end{eqnarray}
\\
\textbf{Physical current density ${\bf{j}} ({\bf r})$ and its
components}
\begin{eqnarray}
&{\bf{j}} ({\bf r}) = {\bf{j}}_p({\bf r}) + {\bf{j}}_d({\bf r}) +
{\bf{j}}_m({\bf r}),&
\end{eqnarray}
where ${\bf{j}}_p({\bf r}), {\bf{j}}_d({\bf r})$, and
${\bf{j}}_m({\bf r})$ are the paramagnetic, diamagnetic and
magnetization current density components, respectively. The
components are
\begin{eqnarray}
&{\bf{j}}_p({\bf r}) = j_{p} (r) ~ \hat{\bf{i}}_{\theta} = \bigg\{
\frac{2 \pi} {\Omega} N^{2} e^{-2 \Omega r^{2}} \nonumber
\\
&\frac{\partial} {\partial r} \bigg[ \int^{\infty}_{0} e^{-\Omega
x^{2}} \frac{1}{x} \bigg(x + c_{2} x^{2} + c_{3} x^{3} + c_{4} x^{4}
\bigg)^{2} I_{0} (2 \Omega r x ) d x \bigg] \bigg\} ~
\hat{\bf{i}}_{\theta} \nonumber \\
&= \frac{\pi N^2 e^{- 3  \Omega r^2/2}}{4 \Omega^4} \bigg\{ 8
e^{\Omega r^2/2} \bigg[ \bigg( 24 C^2+ (6B^2+12 AC ) \Omega
+(2A^2+4B) \Omega^2 + \Omega^3 \bigg) r+
\nonumber\\
&\bigg(36 C^2 \Omega +(6B^2 + 12 AC) \Omega^2 +(A^2+2B)\Omega^3
\bigg) r^3+\bigg(12C^2 \Omega^2 +(B^2 + 2 AC) \Omega^3 \bigg) r^5 +
\nonumber\\
&C^2 \Omega^3 r^7 + \sqrt{\pi \Omega}\bigg[ \bigg(105 BC+30(AB + C)
\Omega+ 12 A \Omega^2 \bigg) r +\bigg( 180 BC \Omega+ 36(AB+
\nonumber\\
& C) \Omega^2++8A\Omega^3 \bigg) r^3 + \bigg( 76 BC  \Omega^2+8
(AB+C) \Omega^3 \bigg) r^5 + 8 BC \Omega^3 r^7 \bigg] I_0(\Omega
r^2/2) +
\nonumber\\
&\sqrt{\pi \Omega}\bigg[ \bigg(15 BC+ 6(AB+C) \Omega + 4 A \Omega^2
\bigg) r+ \bigg( 116 BC \Omega + 28(AB+C) \Omega^2 +
\nonumber\\
& 8 A \Omega^3 \bigg) r^3+ \bigg(68 BC \Omega^2 + 8(AB+C) \Omega^3
\bigg) r^5 + 8 BC  \Omega^3 r^7\bigg] I_1(\Omega r^2/2) \bigg\} ~
\hat{\bf{i}}_{\theta},&
\end{eqnarray}
\begin{eqnarray}
& {\bf{j}}_d({\bf r})= j_{d} (r) ~ \hat{\bf{i}}_{\theta} = r \
\omega_L \ \rho(r) ~ \hat{\bf{i}}_{\theta},&
\end{eqnarray}
\begin{eqnarray}
& {\bf{j}}_m({\bf r})=  j_{m} (r) ~ \hat{\bf{i}}_{\theta} = -
\frac{1}{2} \ \frac{\partial \rho (r)} {\partial r} ~
\hat{\bf{i}}_{\theta}.&
\end{eqnarray}
\\
The asymptotic structures of ${\bf{j}}(\bf{r})$, and its components
are
\begin{eqnarray}
&{\bf{j}}({\bf r}) \ \substack{ \  \\  \\  \sim \\{r \to \ 0}} \
0.0267 \ r - 0.00501 \ r^3 - 0.0000511 \ r^5,
\nonumber\\
& {\bf{j}}({\bf r}) \ \substack{ \  \\  \\  \sim \\{r \to \infty}} \
e^{- \Omega r^2} 10^{-6} \ (1.10 \ r^9 + 8.17 \ r^8 + 25.6 \ r^7),
\nonumber\\
&{\bf{j}}_p({\bf r}) \ \substack{ \  \\  \\  \sim \\{r \to \ 0}} \
0.0149 \ r - 0.00485 \ r^3 + 0.000848 \ r^5,
\nonumber\\
& {\bf{j}}_p({\bf r}) \ \substack{ \  \\  \\  \sim \\{r \to \infty}}
\ e^{- \Omega r^2} 10^{-3} \ (0.00298 \ r^7 + 0.0222 \ r^6 + 0.049 \
r^5 - 0.118 \ r^4),
\nonumber\\
&{\bf{j}}_d({\bf r}) \ \substack{ \  \\  \\  \sim \\{r \to \ 0}} \
10^{-3} \ (5.55 \ r - 0.625 \ r^3 - 0.0230 \ r^5),
\nonumber\\
& {\bf{j}}_d({\bf r}) \ \substack{ \  \\  \\  \sim \\{r \to \infty}}
\ e^{- \Omega r^2} 10^{-6} \ (0.298 \ r^9 + 2.22 \ r^8 + 9.35 \ r^7
+ 17.0 \ r^6),
\nonumber\\
&{\bf{j}}_m({\bf r}) \ \substack{ \  \\  \\  \sim \\{r \to \ 0}} \
10^{-3} \ (6.25 \ r + 0.461 \ r^3 - 0.876 \ r^5),
\nonumber\\
& {\bf{j}}_m({\bf r}) \ \substack{ \  \\  \\  \sim \\{r \to \infty}}
\ e^{- \Omega r^2} 10^{-5} \ (0.0800 \ r^9 + 0.596 \ r^8 + 1.32 \
r^7).
\end{eqnarray}
\\
\textbf{Pair-correlation density $g({\bf rr}')$}
\begin{eqnarray}
g({\bf rr'}) = \frac{2 \ N^2 \ e^{-\Omega (r^2+{r'}\ ^2)} \
\bigg[|{\bf r}- {\bf r'}|+ A \ |{\bf r}- {\bf r'}|^2+ B |{\bf r}-
{\bf r'}|^3+ C |{\bf r}-{\bf r'}|^4 \bigg]^2}{\rho ({\bf r})}.
\end{eqnarray}
\\

\textbf {Single-particle density matrix $\gamma({\bf rr'})$}
\begin{eqnarray}
\gamma ({\bf rr'}) = 2 \ N^2 \ e^{-\Omega (r^2+{r'}\ ^2) /2} \int
e^{- \Omega \ y^2} \ \bigg[|{\bf y}-{\bf r}|+ A |{\bf y}-{\bf r}|^2+
\nonumber\\
B |{\bf y}-{\bf r} \big|^3 + C |{\bf y}-{\bf r} \big|^4 \bigg]
\bigg[|{\bf y}-{\bf r'}|+ A |{\bf y}-{\bf r'}|^2+ B |{\bf y}-{\bf
r'} \big|^3+ C |{\bf y}-{\bf r'} \big|^4 \bigg] \ d {\bf y}.
\end{eqnarray}
\\

\textbf {Electron-interaction field $\boldsymbol{\cal{E}}_ {ee}
({\bf r})$}
\begin{eqnarray}
&\boldsymbol{\cal{E}}_ {ee} ({\bf r}) = \frac{4 \ \pi \ N^2 }{32 \
\Omega^{7/2} \rho (\bf{r})} \bigg\{\sqrt{\pi} \ e^{- 3 \Omega r^2/2}
\bigg[ I_1\big( \Omega\ r^2 /2 \big) \ \bigg( \big[ 15 C^2+ \big(6
B^2+ 12 A C \big) \Omega +
\nonumber\\
&\big( 4 A^2 + 8 B \big) \Omega^2 + 8 \Omega^3 \big] r + \big[ 116
C^2 \  \Omega+ \big( 28 B^2+ 56 A C \big) \ \Omega^2 + \big( 8 A^2 +
\nonumber\\
 &16 B \big)\ \Omega^3 \  \big] r^3  + \big[ 68 C^2 \ \Omega^2 + \big(8 B^2 + 16 A C \big) \ \Omega^3 \big] r^5 + 8 C^2 \ \Omega^3\ r^7 \bigg) +
\nonumber\\
&I_0\big( \Omega\ r^2 /2 \big) \ \bigg( \big[105 C^2 + \big( 30 B^2
+ 60 A C \big) \ \Omega + \big(12 A^2 + 24 AB \big) \ \Omega^2 +
\nonumber\\
&8 \ \Omega^3 \big] r + \big[180 C^2 \ \Omega + \big( 36 B^2 + 72 AC
\big) \ \Omega^2 + \big( 8 A^2 + 16 B \big) \ \Omega^3 \big] r^3  +
\nonumber\\
&\big[76 C^2 \ \Omega^2 + \big(8 B^2 + 16 AC \big) \ \Omega^3 \big]
r^5 + 8 C^2 \ \Omega^3 r^7 \bigg) \ \bigg] + 32 \sqrt{\Omega} \ e^{-
\Omega r^2} \bigg[ \ \bigg(6 BC +
\nonumber\\
 &\big (2 AB + 2 C \big) \  \Omega + A \ \Omega^2 \bigg) r + \bigg( 6 B C \ \Omega + \big(A B + C \big) \ \Omega^2 \bigg) r^3 + BC \ \Omega^2 r^5 \bigg] \ \bigg\},&
\end{eqnarray}
\\
The asymptotic structure of $\boldsymbol{\cal{E}}_{\mathrm{ee}}
(\bf{r})$ and its Hartree ${\boldsymbol{\cal{E}}}_{H} ({\bf{r}})$,
and Pauli-Coulomb ${\boldsymbol{\cal{E}}}_{xc} ({\bf{r}})$
components is
\begin{eqnarray}
&{\cal E}_{ee} (r)& \ \substack{ \  \\  \\  \sim \\{r \to \ 0}} \
0.137 \ r - 0.0360 \ r^3, \ \  \   \  \  \  \ \ {\cal E}_{ee} (r) \
\substack{ \  \\  \\  \sim \\{r \to \infty}} \  \frac{1} {r^2} -
\frac{0.0754} {r^3} - \frac{24.0} {r^4},
\nonumber\\
&{\cal E}_{H} (r)& \ \substack{ \  \\  \\  \sim \\{r \to \infty}} \
\frac{2} {r^2} - \frac{0.0287} {r^3} + \frac{16.3} {r^4}, \  \  \  \
\  \  \  \ {\cal E}_{xc} (r) \ \substack{ \  \\  \\  \sim \\{r \to
\infty}} \ - \frac{1} {r^2} -  \frac{0.0467} {r^3}  - \frac{40.3}
{r^4}.
\end{eqnarray}
\\
\textbf {Electron-interaction energy $E_{ee}$}
\begin{eqnarray}
&E_{ee}= \pi^2 N^2 \bigg\{ \frac{4}{\Omega^5} \big[ 24 BC + 4 \big(
AB + C \big) \Omega +A \ \Omega^2 \big] + \big[ 105 C^2 +
\nonumber\\
& 15 \big(B^2 + 2 AC) \ \Omega + 3 \big(A^2 +2 AB) \Omega^2 + \
\Omega^3 \big] \bigg\} = 0.254158 \ (a.u.)^{\star}.&
\end{eqnarray}
\\

\textbf{Kinetic energy tensor $t_{\alpha \beta} [{\bf r; \gamma}]$ }
\begin{eqnarray}
t_{\alpha \beta} [{\bf r; \gamma}]= \frac{r_\alpha \ r_\beta} {r^2}
\ f(r) +\delta_{\alpha \beta} \ k(r),
\end{eqnarray}
where
\begin{eqnarray}
f(r) = \pi \ N^2 e^{- 2 \Omega r^2} \bigg[\frac{r}{\Omega}
\frac{\partial f_1(r)} {\partial r} - 2 \ r \frac{\partial f_2(r)}
{\partial r} \bigg] + \frac{\Omega^2} {2} \ r^2 \rho(r),
\end{eqnarray}
\begin{eqnarray}
k(r) = \pi \ N^2 \ e^{- 2 \Omega r^2} \bigg[ \frac{f_1(r)} {\Omega}
+ 2 f_3(r) \bigg],
\end{eqnarray}
\begin{eqnarray}
&f_1(r) = \frac{e^{\Omega r^2 / 2}} {8 \ \Omega^3} \bigg\{ 4 \
e^{\Omega^2 r^2/2} \bigg[ \ \bigg(90 C^2 \ \Omega + \big(8B^2 + 14
AC \big) \Omega^2 \bigg) \big(1 + r^2 \big)+ 15 C^2 \ \Omega^2 r^4
\bigg] +
\nonumber\\
& \sqrt {\pi \Omega} \bigg[\bigg(165 BC + \big(30 AB + 18 C \big)
\Omega + 4 A \ \Omega^2 \bigg) + r^2 \bigg( 198 BC \ \Omega + \big(
20 AB +
\nonumber\\
&12 C \big) \Omega^2 \bigg) + 44 BC \Omega^2 r^4 \bigg] I_0(\Omega
r^2/2) + \sqrt{\pi \Omega} \  \bigg[ \bigg(33 BC + \big(10 AB + 6 C
\big) \Omega +
\nonumber\\
 &4 A \Omega^2 \bigg) + r^2 \bigg( 154 BC \Omega + \big(20 AB + 12 C \big) \Omega^2 \bigg) + 44 BC \Omega^2 r^4 \bigg] I_1(\Omega r^2/2) \bigg\}&
\end{eqnarray}
\begin{eqnarray}
&f_2(r) = \frac{e^{\Omega r^2/2}}{16 \Omega^4} \bigg\{ 8 e^{\Omega
r^2/2} \bigg[ \bigg(24 C^2 + \big(6 B^2 + 12 AC \big) \Omega +
\big(2 A^2+ 4B \big) \Omega^2 + \Omega^3 \bigg)+
\nonumber\\
&\bigg( 72 C^2 \Omega + \big(12 B^2 + 24 AC \big) \Omega^2 + \big(2
A^2 + 4 B \big) \Omega^3 \bigg) r^2 + \bigg( 36 C^2 \Omega^2+ \big(3
B^2 +
\nonumber\\
&6AC \big) \Omega^3 \bigg) r^4 +4 C^2 \Omega^3 r^6 \bigg] +\sqrt{\pi
\Omega} \bigg[ \bigg( 105 BC+ 30 \big(AB + C) \Omega + 12 A \Omega^2
\bigg) +
\nonumber\\
&28 BC \Omega^3 r^6+ \bigg(315 BC \Omega +60 \big( AB + C \big)
\Omega^2 +12 A \Omega^3 \bigg) r^2 + \bigg( 196 BC \Omega^2 +
\nonumber\\
&20 \big(AB+ C \big) \Omega^3 \bigg) r^4 \bigg] I_0(\Omega r^2/2) +
\sqrt{\pi \Omega} \ \Omega \bigg[ \bigg( 161 BC + 40 \big(AB + C
\big) \Omega +
\nonumber\\
& 12 A \Omega^2 \bigg) r^2 + \bigg( 168 BC \Omega + 20 \big(AB + C
\big) \Omega^2 \bigg) r^4 + 28 BC \Omega^2 r^6 \bigg] I_1(\Omega
r^2/2) \bigg\}&
\end{eqnarray}
\begin{eqnarray}
&f_3(r) = \frac{e^{\Omega r^2/2}}{8 \Omega^4} \bigg\{ 4 e^{\Omega
r^2/2} \bigg[ \bigg(6 C^2+ 2 \big( 2B^2 +4AC \big) \Omega + (A^2 + 2
B \big) \Omega^2 + \Omega^3 \bigg) +
\nonumber\\
& \bigg(18 C^2 \Omega + 4 (B^2 + 2AC) \Omega^2 +(A^2+2B) \Omega^3
\bigg) r^2 + \bigg(9 C^2 \Omega^2 +(B^2+
\nonumber\\
&2AC) \Omega^3 \bigg) r^4 + C^2 \Omega^3 r^6 \bigg] + \sqrt{\pi
\Omega} \bigg[ \bigg( 15BC + 6(AB +C) \Omega+4 A \Omega^2 \bigg) +
\nonumber\\
&\bigg( 45 BC \Omega + 12 (AB+C) \Omega^2 +4A \Omega^3 \bigg) r^2 +
\bigg( 28 BC \Omega^2 +4 (AB +C) \Omega^3 \bigg) r^4 +
\nonumber\\
& 4BC \Omega^3 r^6 \bigg] I_0(\Omega r^2/2) + \sqrt{\pi \Omega} \
\Omega \bigg[ \bigg( 23BC + 8(AB+C) \Omega+ 4 A \Omega^2 \bigg) r^2
+
\nonumber\\
&\bigg(24BC \Omega + 4(AB+C) \Omega^2 \bigg) r^4 + 4BC \Omega^2 r^6
\bigg] I_1(\Omega r^2/2) \bigg \}.&
\end{eqnarray}
\\

\textbf{Kinetic `force' $z_{\alpha} [\bf{r}; \gamma]$}
\begin{eqnarray}
z_{\alpha} [{\bf r; \gamma}]= \frac{2 \ r_\alpha} {r} \bigg \{
\frac{\partial [f(r) + k(r)]} {\partial r} + \frac{f(r)} {r} \bigg
\},
\end{eqnarray}
where the functions $f(r)$ and $k(r)$ are given in Eq. (A20) and
(A21), respectively.  The asymptotic structures are
\begin{eqnarray}
z(r) \ \substack{ \  \\  \\  \sim \\{r \to 0}} \ 0.00174 \ r +
0.00811 \ r^3 + \ldots
\end{eqnarray}
\begin{eqnarray}
z(r) \ \substack{ \  \\  \\  \sim \\{r \to \infty}} \   e^{- \Omega
r^2} 10^{-5} (- 0.0116 r^{11} - 0.0860 r^{10} + 0.218 r^9 - 0.0700
r^8 + 5.55 r^7
 + \ldots) .
\end{eqnarray}
\\

\textbf {Kinetic Energy $T$}
\begin{eqnarray}
&T = \frac{N_R^2 \ \pi}{4} + \frac{N_r^2 \ \pi} {\Omega^4}\bigg(
480C^2 + 12 A^2 \Omega^2 + 3 A \sqrt{\pi /2} \ \Omega^{3/2} (13 B+ 3
\Omega) + 1.5 C (85 B \sqrt{2 \pi \Omega} +
\nonumber\\
&  64 A \Omega + 5 \sqrt{2 \pi} \Omega^{3/2}) + 4\Omega (16 B^2 +4 B
\Omega + \Omega^2) \bigg) = 0.615577~ (a.u.)^{\star},&
\end{eqnarray}
where $N_r = 0.05431655771, N_R = 0.4136182782$.
\\

\textbf{Lorentz ${\boldsymbol{\cal{L}}} ({\bf{r}})$ and Internal
Magnetic ${\boldsymbol{\cal{I}}}_{m} ({\bf{r}})$ Fields}

\begin{equation}
{\boldsymbol{\cal{L}}} ({\bf{r}}) = \frac{2 \omega_{L} j (r)} {\rho
(r)} ~ \hat{\bf{i}}_{r},
\end{equation}
\begin{equation}
 {\boldsymbol{\cal{I}}}_{m} ({\bf{r}}) = \bigg[ - \frac{2
\omega_{L} j (r)} {\rho (r)} + \omega^{2}_{L} r \bigg] ~
\hat{\bf{i}}_{r},
\end{equation}
\begin{equation}
{\boldsymbol{\cal{M}}} ({\bf{r}}) = - [{\boldsymbol{\cal{L}}}
({\bf{r}})  + {\boldsymbol{\cal{I}}}_{m} ({\bf{r}})] = -
\omega^{2}_{L} r  ~ {\bf{i}}_{r},
\end{equation}
where $j (r) = j_{p} (r) + j_{d} (r) + j_{m} (r)$ (see Eqs.
(A10-A12).)
\\

\textbf {External Electrostatic $E_{es}$ and Magnetostatic $E_{mag}$
Energies}
\begin{eqnarray}
E_{es} + E_{mag} &=& \int \rho (r) \bigg[ \frac{1}{2} k_\mathrm{eff}
r^{2} \bigg] d {\bf{r}} \nonumber \\
&=& k_\mathrm{eff}
\frac{\pi^{2} N^{2}} {4 \Omega^{7}} \bigg( 64 \Omega^2 \ (A^2 + 2 B)
+ 480 \Omega (2 AC + B^2) + 135 \sqrt{2 \pi} \ \Omega^{3/2}(AB+C)
\nonumber \\
&+& \sqrt{2 \pi \Omega} \ (21 A \Omega^2 + 1155 BC) + 4608 C^2 + 12
\Omega^3 \bigg) \nonumber \\
&=& 0.742657~(a.u.)^{\star}
\end{eqnarray}

\textbf {Expectation values}

With the complete elliptical integrals \cite{23}
\begin{eqnarray}
K (p) &=& \int^{\pi/2}_{0} \frac{d \theta} {\sqrt{1-p^{2} \sin^{2}
\theta}}, \\
E (p) &=& \int^{\pi/2}_{0} \sqrt{1-p^{2} \sin^{2} \theta} d \theta,
\end{eqnarray}
the expressions and values of the various expectations are:
\begin{eqnarray}
&<r> = \int \rho(r) \ r \ d {\bf r}
\nonumber\\
&=\frac{\pi^2  N^2}{2 \Omega^{13/2}} \bigg[\frac{47}{2} \sqrt{\pi}\
\Omega^2 (A^2+2B) +\frac{639}{4} \sqrt{\pi} \ \Omega (2AC+B^2) +
\sqrt{2}\  \Omega^{3/2} \bigg( 174 E(1/2) -
\nonumber\\
& 47 K(1/2) \bigg) (AB+C) + 2 \sqrt{2}\  A \Omega^{5/2} \bigg( 15
E(1/2) - 4 K(1/2) \bigg) + 5 \sqrt{2 \Omega}\  BC
\nonumber\\
&\bigg(273 E(1/2) - 74 K(1/2) \bigg) + \frac{11313}{8} \sqrt{\pi} \
C^2 +5 \sqrt{\pi} \ \Omega^3 \bigg] = 5.823553 \ (a.u.)^{\star},&
\end{eqnarray}
\begin{eqnarray}
&<r^2> = \int \rho(r) \ r^2 \ d {\bf r}
\nonumber\\
&=\frac{\pi^2  N^2}{2 \Omega^7} \bigg( 4608 C^2 + 1155 BC \sqrt{2
\pi \Omega} + 480 (B^2 + 2 AC) \Omega +135 (AB +
\nonumber\\
& C) \sqrt{2 \pi} \Omega^{3/2} +64 (A^2+2B) \Omega^2 +21 A \sqrt{2
\pi} \  \Omega^{5/2} + 12 \Omega^3 \bigg) = 20.567403 \
(a.u.)^{\star},&
\end{eqnarray}
\begin{eqnarray}
&\bigg< \frac{1} {r} \bigg> = \int \rho(r) \ \bigg(\frac{1} {r}
\bigg) \ d {\bf r}
\nonumber\\
&=\frac{\pi^2 N^2}{8 \Omega^{11/2}} \bigg[ 76 \sqrt{\pi}\ \Omega^2
(A^2 + 2 B) + 378 \sqrt{\pi} \ \Omega (2 AC +B^2) + 48 \sqrt{2} \
\Omega^{3/2} \bigg( 9 E(1/2) - 2 K(1/2) \bigg)
\nonumber\\
&  (AB+C) +16 \sqrt{2} \ A \Omega^{5/2} \bigg( 6 E(1/2) - K(1/2)
\bigg) + 8 \sqrt{2 \Omega} \ B C \bigg( 336 E(1/2) - 83 K(1/2)
\bigg) +
\nonumber\\
& 2601 \sqrt{\pi} \ C^2 + 24 \sqrt{\pi} \  \Omega^3 \bigg] =
1.041717 \ (a.u.)^{\star},&
\end{eqnarray}
\begin{eqnarray}
&<\delta ({\bf r})> = \int \rho (r) \delta ({\bf{r}}) d {\bf{r}} =
\rho(0)
\nonumber\\
&=\frac{N^2 \pi}{4 \Omega^5} \bigg[ 3 \sqrt{\pi\Omega} \ \bigg( 35
BC +10 (AB+C) \Omega + 4 A \Omega^2 \bigg) + 8 \bigg(24 C^2
+6(B^2+2AC) \Omega +
\nonumber\\
&2(A^2 + 2B) \Omega^2 +\Omega^3 \bigg)  \ \bigg] = 0.0555377 \
(a.u.)^{\star},&
\end{eqnarray}

\section{Derivation of the Kinetic-Energy Tensor and Kinetic
`Force' for the $2^{3} S$ State of the Quantum Dot}

The spatial part $\Psi ({\bf{r}}_{1} {\bf{r}}_{2})$ of the first
excited triplet state wave function is
\begin{equation}
\Psi ({\bf{r}}_{1},  {\bf{r}}_{2}) = Ne^{i \theta} e^{- \Omega
(r^{2}_{1} + r^{2}_{2})/2} g_{0} (u),
\end{equation}
\begin{equation}
g_{0} (u) = u + c_{2} u^{2} + c_{3} u^{3} + c_{4} u^{4},
\end{equation}
where the values of the coefficients $N, \Omega, c_{2}, c_{3},
c_{4}$ are given in Sect. II, and $\theta$ is the angle of the
relative coordinate ${\bf{u}} = {\bf{r}}_{2} - {\bf{r}}_{1}$.

The kinetic energy tensor $t_{\alpha \beta} ({\bf{r}} ; \gamma)$ is
defined as
\begin{equation}
t_{\alpha \beta} ({\bf{r}} ; \gamma) = \frac{1}{4} \bigg[
\frac{\partial^{2}} {\partial r_{p \alpha} \partial r_{q \beta}} +
\frac{\partial^{2}} {\partial r_{p \beta} \partial r_{q \alpha}}
\bigg] \gamma ({\bf{r}}_{p}, {\bf{r}}_{q}) \bigg|_{{\bf{r}}_{p} =
{\bf{r}}_{q} = {\bf{r}}},
\end{equation}
where the single-particle density matrix $\gamma ({\bf{r}}_{p}
{\bf{r}}_{q})$ is
\begin{equation}
\gamma ({\bf{r}}_{p}, {\bf{r}}_{q}) = 2 \int \Psi^{\star}
({\bf{r}}_{p}, {\bf{r}}_{2}) \Psi^{\star} ({\bf{r}}_{q},
{\bf{r}}_{2}) d {\bf{r}}_{2}.
\end{equation}
Hence, the components of the tensor are
\begin{eqnarray}
&t_{xx} = \int \bigg( \frac{\partial \Psi^{\star}_{p,2}} {\partial
x_{p}} ~ \frac{\partial \Psi_{q,2}} {\partial x_{q}} \bigg)
\bigg|_{{\bf{r}}_{p} = {\bf{r}}_{q} = r} d {\bf{r}}_{2}, \\
& t_{xy} = \frac{1}{2} \int \bigg( \frac{\partial
\Psi^{\star}_{p,2}} {\partial x_{p}} ~ \frac{\partial \Psi_{q,2}}
{\partial y_{q}} + \frac{\partial \Psi^{\star}_{p,2}} {\partial
y_{p}} ~ \frac{\partial \Psi_{q,2}} {\partial x_{q}} \bigg)
\bigg|_{{\bf{r}}_{p} = {\bf{r}}_{q} = {\bf{r}}} d {\bf{r}}_{2}, \\
& t_{y x} = t_{x y} , \\
&t_{y y} = \int \bigg(\frac{\partial \Psi^{\star}_{p,2}} {\partial
y_{p}} ~ \frac{\partial \Psi_{q,2}} {\partial y_{q}} \bigg)
\bigg|_{{\bf{r}}_{p} = {\bf{r}}_{q} = {\bf{r}}} d {\bf{r}}_{2}.
\end{eqnarray}
We next determine the derivatives in the components of the tensor.
\\

\emph{(i)} Writing ${\bf{r}}_{1} = {\bf{r}}_{p}$,
\begin{equation}
\frac{\partial} {\partial x_{p}} e^{- \Omega (r^{2}_{p} +
r^{2}_{2})/2} \bigg|_{{\bf{r}}_{p} = r} = - \Omega x e^{-\Omega
(r^{2} + r^{2}_{2})/2} .
\end{equation}
\emph{(ii)} Writing ${\bf{r}}_{p} = {\bf{r}}$, and defining
${\bf{r}}_{2} - {\bf{r}} = {\bf{r}}_{3}$,
\begin{eqnarray}
&\frac{\partial} {\partial x_{p}} | {\bf{r}}_{2} - {\bf{r}}_{p}|
= - \frac{x_{3}} {r_{3}} , \\
&\frac{\partial} {\partial x_{p}} | {\bf{r}}_{2} - {\bf{r}}_{p}|^{2}
= - 2x_{3} , \\
&\frac{\partial} {\partial x_{p}} | {\bf{r}}_{2} - {\bf{r}}_{p}|^{3}
= - 3r_{3} x_{3} , \\
&\frac{\partial} {\partial x_{p}} | {\bf{r}}_{2} - {\bf{r}}_{p}|^{4}
= - 4r_{3}^{2} x_{3}. &
\end{eqnarray}
Thus,
\begin{eqnarray}
&\frac{\partial} {\partial x_{p}} g_{0} (r_{3}) = - x_{3}  \big(
\frac{1}{r_{3}} +
2c_{2} + 3 c_{3} r_{3} + 4c_{4} r^{2}_{3} \big ) \\
& = - x_{3} g_{1} (r_{3}).&
\end{eqnarray}
\emph{(iii)}
With

\begin{equation}
\theta_{p,2} = \tan^{-1} \big (\frac{y_{2} - y_{p}} {x_{2} - x_{p}}
\big),
\end{equation}
\begin{equation}
\frac{\partial} {\partial x_{p}} e^{- i \theta_{p,2}} = - i
\frac{y_{3}} {r^{2}_{3}} e^{- i \theta_{p,2}}.
\end{equation}
Hence, the first derivative of the integrand of $t_{xx}$ of (B5) is
\begin{equation}
\frac{\partial \Psi^{\star}_{p,2}} {\partial x_{p}} = Ne^{-
\frac{\Omega} {2} (2 r^{2} + r^{2}_{3} + 2 {\bf{r}} \cdot
{\bf{r}}_{3})} \big [ - \Omega x g_{0} - x_{3} g_{1} - i
\frac{y_{3}} {r^{2}_{3}} g_{0} \big] e^{- i \theta_{3}} .
\end{equation}
In a similar manner, the second derivative $\partial
\Psi_{q,2}/\partial x_{q}$ is obtained, so that the integrand of
$t_{xx}$ of (B5) is
\begin{equation}
\frac{\partial \Psi^{\star}_{p,2}} {\partial x_{p}} \cdot
\frac{\partial \Psi_{q,2}} {\partial x_{q}} = N^{2} e^{- \Omega (2
r^{2} + r^{2}_{3} + 2 {\bf{r}} \cdot {\bf{r}}_{3})} \bigg
[\Omega^{2} x^{2} g^{2}_{0} + x^{2}_{3} g^{2}_{1} + 2 \Omega x x_{3}
g_{0} g_{1} + \frac{y^{2}_{3}} {r^{4}_{3}} g^{2}_{0} \bigg ].
\end{equation}
Similarly, the integrand of $t_{yy}$ of (B8) is
\begin{equation}
\frac{\partial \Psi^{\star}_{p,2}} {\partial y_{p}} \cdot
\frac{\partial \Psi_{q,2}} {\partial y_{p}} = N^{2} e^{- \Omega (2
r^{2} + r^{2}_{3} + 2 {\bf{r}} \cdot {\bf{r}}_{3})} \bigg
[\Omega^{2} y^{2} g^{2}_{0} + y^{2}_{3} g^{2}_{1} + 2 \Omega y y_{3}
g_{0} g_{1} + \frac{x^{2}_{3}} {r^{4}_{3}} g^{2}_{0} \bigg ],
\end{equation}
and that of $t_{xy}$ of (B6) is
\begin{eqnarray}
&\frac{1}{2} \bigg(\frac{\partial \Psi^{\star}_{p,2}} {\partial
x_{p}} \cdot \frac{\partial \Psi_{q,2}} {\partial y_{q}} +
\frac{\partial \Psi^{\star}_{p,2}} {\partial y_{p}} \cdot
\frac{\partial \Psi_{q,2}} {\partial x_{q}} \bigg) = N^{2} e^{-
\Omega (2 r^{2} + r^{2}_{3} + 2 {\bf{r}} \cdot
{\bf{r}}_{3})} ~~~~~~~~~~~~~~~~~~ \nonumber \\
& \bigg [\Omega^{2} xy g^{2}_{0} + x_{3}y_{3} g^{2}_{1} + \Omega x
y_{3} g_{0} g_{1} + \Omega y x_{3} g_{0} g_{1} - \frac{x_{3} y_{3}}
{r^{4}_{3}} g^{2}_{0} \bigg ].&
\end{eqnarray}

Let us first consider the off-diagonal component $t_{xy}$.  In this
component, consider the contribution of the first term of (B21) in
the square parentheses which is
\begin{eqnarray}
& N^{2} e^{- 2 \Omega r^{2}} xy \Omega^{2} \int g^{2}_{0} (r_{3})
e^{- \Omega (r^{2}_{3} + 2 {\bf{r}} \cdot {\bf{r}}_{3})} d
{\bf{r}}_{3} \\
& = (xy \Omega^{2}) \bigg[ N^{2} e^{-2 \Omega r^{2}}
\int^{\infty}_{0} e^{-\Omega r^{2}_{3}} g^{2}_{0} (r_{3}) r_{3} d
r_{3} \int^{2 \pi}_{0} e^{-2 \Omega r r_{3} \cos \theta_{3}} d
\theta_{3} \bigg]
\\
& = (xy \Omega^{2}) \bigg[ N^{2} e^{-2 \Omega r^{2}} 2 \pi
\int^{\infty}_{0} e^{-\Omega r^{2}_{3}} g^{2}_{0} (r_{3}) I_{0} (2
\Omega r r_{3}) r_{3} d r_{3} \bigg] , \\
& = \frac{(x y \Omega^{2})} {2} \rho (r),&
\end{eqnarray}
where $I_{0} (x)$ is the zeroth-order modified Bessel function
\cite{23}.  (This term can be written more generally as $(r_{\alpha}
r_{\beta} \Omega^{2} \rho (r)/2)$, where $r_{\alpha}, r_{\beta}$
represent either $x$ or $y$.)

The vector components $x_{3}$ and $y_{3}$ of the second term in the
square parentheses of (B21) can be eliminated through the equalities
\begin{equation}
x_{3} e^{-2 \Omega {\bf{r}} \cdot {\bf{r}}_{3}} = - \frac{1}{2
\Omega} \frac{\partial} {\partial x} e^{-2 \Omega {\bf{r}}  \cdot
{\bf{r}}_{3}},
\end{equation}
and
\begin{equation}
y_{3} e^{-2 \Omega {\bf{r}} \cdot {\bf{r}}_{3}} = - \frac{1}{2
\Omega} \frac{\partial} {\partial y} e^{-2 \Omega {\bf{r}}  \cdot
{\bf{r}}_{3}},
\end{equation}
and then by evaluating the $d \theta_{3}$ integral of (B6) first,
the contribution of the second term of (B21) to $t_{xy}$ is
\begin{equation}
\frac{2 \pi N^{2}} {4 \Omega^{2}} e^{- 2 \Omega r^{2}}
\frac{\partial^{2}} {\partial x \partial y} \int^{\infty}_{0} r_{3}
g^{2}_{1} (r_{3}) e^{- \Omega r^{2}_{3}} I_{0} (2 \Omega r r_{3}) d
r_{3}.
\end{equation}
As the lowest-order of $g_{1} (r_{3})$ is $1/r_{3}$, the integrand
of (B28) goes as $1/r_{3}$, which is singular at $r_{3} =0$.  In
order to eliminate the singularity, we employ
\begin{equation}
\frac{\partial} {\partial y}  I_{0} (2 \Omega r r_{3}) = \frac{2
\Omega y r_{3}} {r} I_{1} (2 \Omega r r_{3}),
\end{equation}
where $I_{1} (x)$ is the first-order modified Bessel function
\cite{23}.

The contribution of the fifth term of (B21) to the integral of (B6)
also goes as $1/r_{3}$ to lowest-order, and the singularity is
treated as above.  Then by evaluating the $d r_{3}$ integral, and
employing the equality for a general function $w (r)$ as follows:
\begin{equation}
\frac{\partial} {\partial r_{\alpha}} \big[ r_{\beta} w (r) \big] =
\delta_{\alpha \beta} w (r) + \frac{r_{\alpha} r_{\beta}} {r}
\frac{\partial w (r)} {\partial r},
\end{equation}
the contribution of the combination of the second and fifth terms of
(B21) to (B6) for $t_{xy}$ may be written as
\begin{equation}
\frac{\pi N^{2}} {\Omega} e^{-2 \Omega r^{2}} \bigg[\delta_{\alpha
\beta} f_{1} (r) + \frac{r_{\alpha} r_{\beta}} {r} \frac{\partial
f_{1} (r)} {\partial r} \bigg ],
\end{equation}
where
\begin{equation}
f_{1} (r) = \frac{1}{r} \int^{\infty}_{0} \big(g^{2}_{1} -
\frac{g^{2}_{0}} {r^{4}_{3}} \big) e^{- \Omega r^{2}_{3}} r^{2}_{3}
I_{1} (2 \Omega r r_{3}) d r_{3}.
\end{equation}
(See (A22) for $f_{1} (r)$).

To evaluate contribution of the third and fourth cross-terms of
(B21) to (B6), which are identical, we apply the equalities (B26)
and (B27), evaluate the $d \theta_{3}$ integral first (no
singularity in this case), then evaluate the $d r_{3}$ integral, and
employ the following equality for any function $w (r)$
\begin{equation}
r_{\beta} \frac{\partial w (r)} {\partial r_{\alpha}} =
\frac{r_{\alpha} r_{\beta}} {r} \frac{\partial w (r)} {\partial r}.
\end{equation}
Then the sum of the cross-terms may be written as
\begin{equation}
- 2 \pi N^{2} e^{-2 \Omega r^{2}} \bigg(\frac{r_{\alpha} r_{\beta}}
{r} \bigg) \frac{\partial f_{2} (r)} {\partial r},
\end{equation}
where
\begin{equation}
f_{2} (r) = \int^{\infty}_{0} r_{3} g_{1} (r_{3}) g_{0} (r_{3}) e^{-
\Omega r^{2}_{3}} I_{0} (2 \Omega r r_{3}) d r_{3}.
\end{equation}
(See (A23) for $f_{2} (r)$).

Next consider the diagonal elements $t_{xx}$ and $t_{yy}$ of (B5)
and (B8), respectively.  The first three terms of the corresponding
integrands given by (B19) and (B20) are evaluated in the same way as
the first 3 terms of the off-diagonal element $t_{xy}$ as described
above.

Note that the contribution of the last term of (B19) to $t_{xx}$ is
proportional to $y^{2}_{3}$ (instead of $x^{2}_{3}$), whereas that
of the last term of (B20) to $t_{yy}$ is proportional to $x^{2}_{3}$
(instead of $y^{2}_{3}$).  Since $y^{2}_{3} = r^{2}_{3} -
x^{2}_{3}$, the last term of (B19) may be written as $y^{2}_{3}
g^{2}_{0}/r^{4}_{3} = (r^{2}_{3} - x^{2}_{3}) g^{2}_{0}/r^{4}_{3}$.
This term may be further generalized to include the corresponding
term of the off-diagonal element $t_{xy}$ by writing it as
\begin{equation}
\big( \delta_{\alpha \beta} r^{2}_{3} - r_{3 \alpha} r_{3 \beta}
\big) \frac{g^{2}_{0}} {r^{4}_{3}}.
\end{equation}
Notice that (B36) is identical to the fifth term in (B21) for
$t_{xy}$, because $\delta_{\alpha \beta} = 0$ when $\alpha \neq
\beta$.  (In this case $\alpha = x$, and $\beta = y$).

We next determine the contribution of the $\delta_{\alpha \beta}
r^{2}_{3}$ term of (B36) to $t_{xx}$. From (B19), this contribution
is
\begin{eqnarray}
& N^{2} e^{- 2 \Omega r^{2}} \int^{\infty}_{0} e^{- \Omega
r^{2}_{3}} \frac{g^{2}_{0}} {r^{4}_{3}} \cdot r^{2}_{3} \cdot r_{3}
d r_{3} \int^{2 \pi}_{0} e^{-2 \Omega r r_{3} \cos \theta_{3}} d
\theta_{3} \\
& = 2 \pi N^{2} e^{-2 \Omega r^{2}} f_{3} (r) &
\end{eqnarray}
where
\begin{equation}
f_{3} (r) = \int^{\infty}_{0} e^{- \Omega r^{2}_{3}}
\frac{g^{2}_{0}} {r_{3}} I_{0} (2 \Omega r r_{3}) d r_{3}.
\end{equation}
(See (A24) for $f_{3} (r)$).

The second term of (B36) is the same as the last term of (B21), and
its contribution has been previously evaluated.

Thus, in summing all the requisite terms, the tensor $t_{\alpha
\beta}$ may be written as
\begin{equation}
t_{\alpha \beta} (r) = \frac{r_{\alpha} r_{\beta}} {r^{2}} f (r) +
\delta_{\alpha \beta} k (r)
\end{equation}
where $f (r)$ and $k (r)$ are defined in (A20) and (A21).

The kinetic `force' component is defined as
\begin{equation}
z_{\alpha} (r) = 2 \sum^{2}_{\beta = 1} \nabla_{\beta} t_{\alpha
\beta} (r).
\end{equation}
Upon substituting $t_{\alpha \beta} (r)$ of (B40) into (B41) we
obtain
\begin{equation}
z_{\alpha} (r) = 2 \sum^{2}_{\beta = 1} \frac{\partial} {\partial
r_{\beta}} \bigg[ \frac{r_{\alpha} r_{\beta}} {r^{2}} f (r) +
\delta_{\alpha \beta} k (r) \bigg],
\end{equation}
where $f (r)$ and $k (r)$ are given in (A20) and (A21).

For the $2D$ coordinate system, it can be shown
\begin{eqnarray}
& \sum^{2}_{\beta = 1} \frac{\partial} {\partial r_{\beta}}
(r_{\alpha} r_{\beta}) = 3 r_{\alpha} , \\
& \sum^{2}_{\beta = 1} r_{\alpha} r_{\beta} \frac{\partial}
{\partial r_{\beta}} \big[ \frac{f (r)} {r^{2}} \big] = r_{\alpha}
\big[ \frac{1} {r} \frac{\partial f (r)} {\partial r} - \frac{2 f
(r)} {r^{2}} \big],
\\
& \sum^{2}_{\beta = 1} \frac{\partial} {\partial r_{\beta}}
\big[\delta_{\alpha \beta} k (r) \big] = \frac{r_{\alpha}} {r}
\frac{\partial k (r)} {\partial r}. &
\end{eqnarray}
Finally, by substituting (B43), (B44), and (B45) into (B42), we
obtain the components of the kinetic `force' as given in (A25).

\newpage

\begin{figure}
\includegraphics[width=0.7\textwidth]{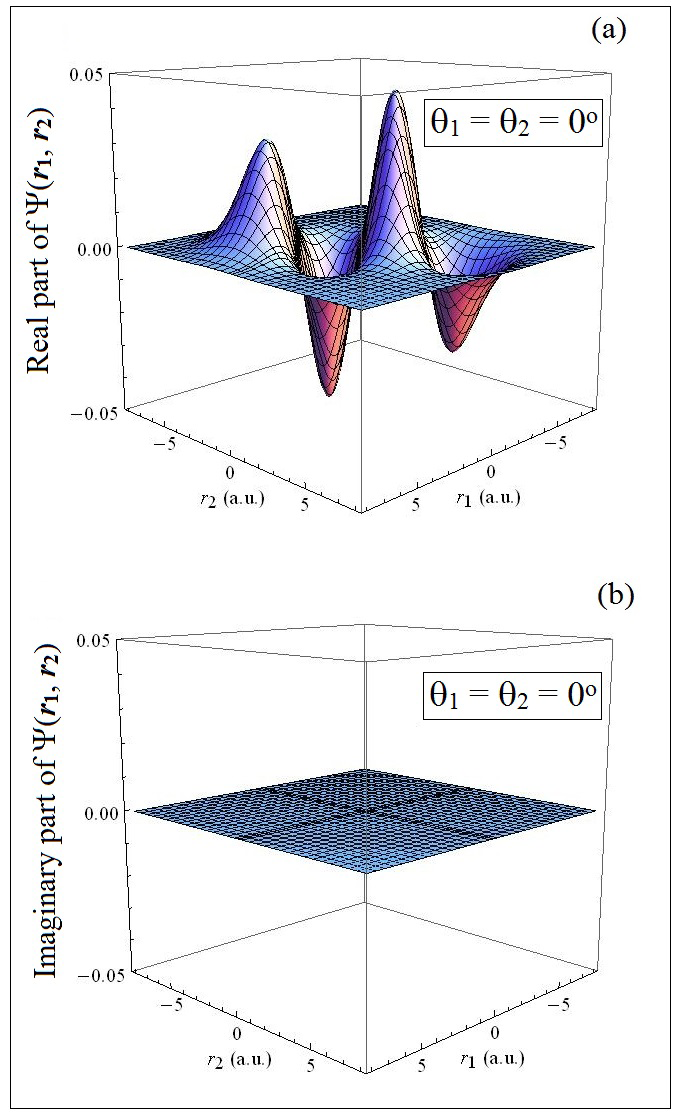}
\caption{(a) Structure of the Real component of the spatial part
$\Psi ({\bf{r}}_{1} {\bf{r}}_{2})$ of the triplet $2^{3} S$ wave
function of the quantum dot in a magnetic field. The angles
$\theta_{1}, \theta_{2}$ of the vectors ${\bf{r}}_{1}$ and
${\bf{r}}_{2}$ are measured from the $+x$-axis.  In this Fig. 1,
these angles are $\theta_{1} = \theta_{2} = 0^{\circ}$, which means
vectors ${\bf{r}}_{1}$ and ${\bf{r}}_{2}$ are oriented along the $x$
axis. (b) The corresponding structure of the Imaginary part of $\Psi
({\bf{r}}_{1} {\bf{r}}_{2})$.}
\end{figure}

\begin{figure}
\includegraphics[width=0.7\textwidth]{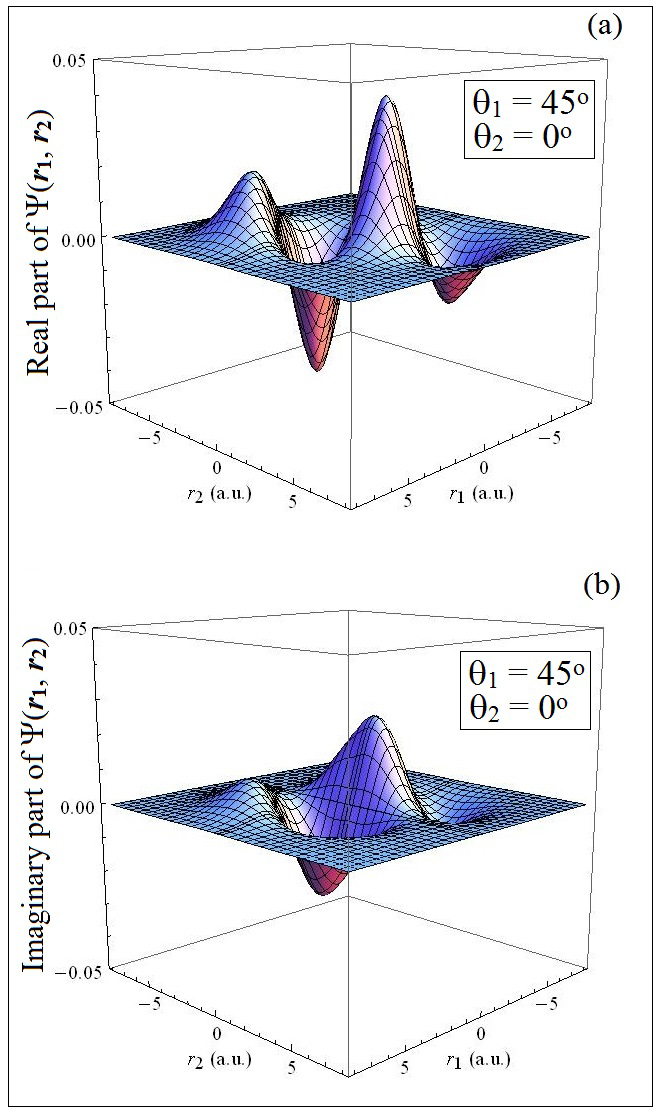}
\caption{Same as in Fig. 1 except that $\theta_{1} = 45^{\circ},
\theta_{2} = 0^{\circ}$.  In this figure, the vector ${\bf{r}}_{2}$
is along the $x$ axis.}
\end{figure}

\begin{figure}
\includegraphics[width=0.7\textwidth]{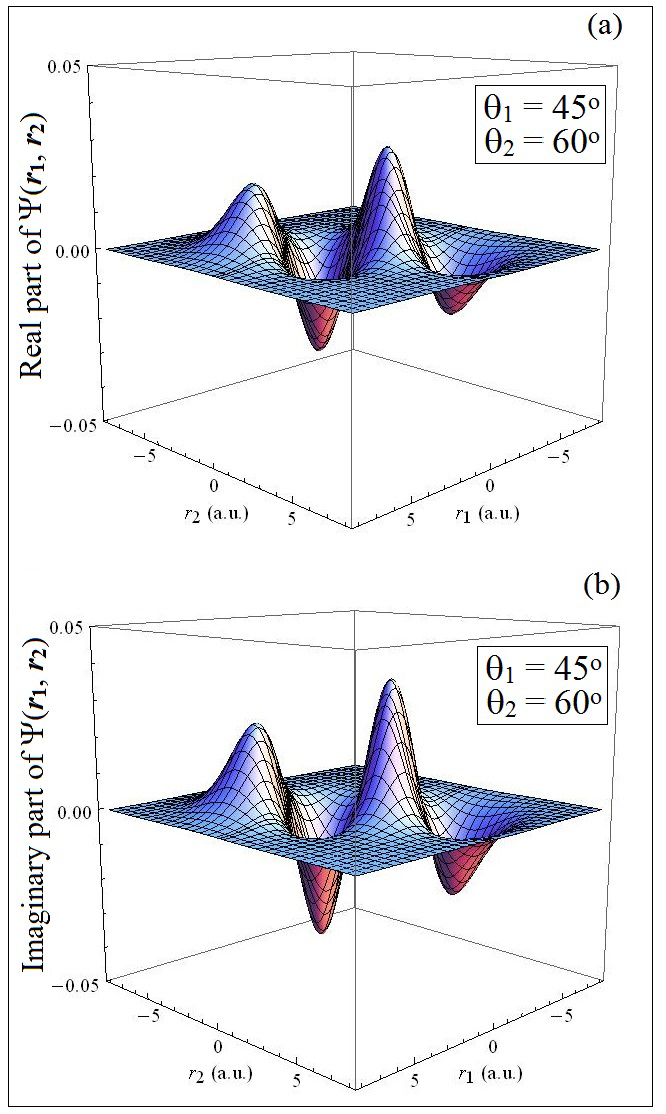}
\caption{Same as in Fig. 1 except that $\theta_{1} = 45^{\circ},
\theta_{2} = 60^{\circ}$.}
\end{figure}

\begin{figure}
\includegraphics[width=0.7\textwidth]{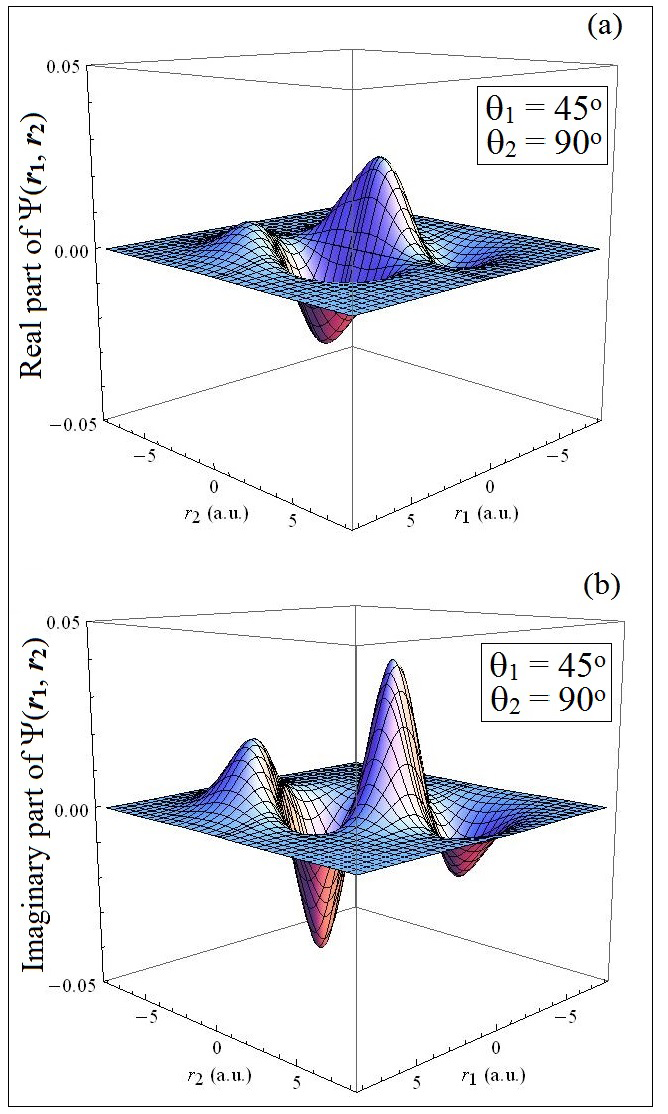}
\caption{Same as in Fig. 1 except that $\theta_{1} = 45^{\circ},
\theta_{2} = 90^{\circ}$.  In this figure, the vector ${\bf{r}}_{2}$
is along the $y$-axis.}
\end{figure}

\begin{figure}
\includegraphics[width=0.8\textwidth]{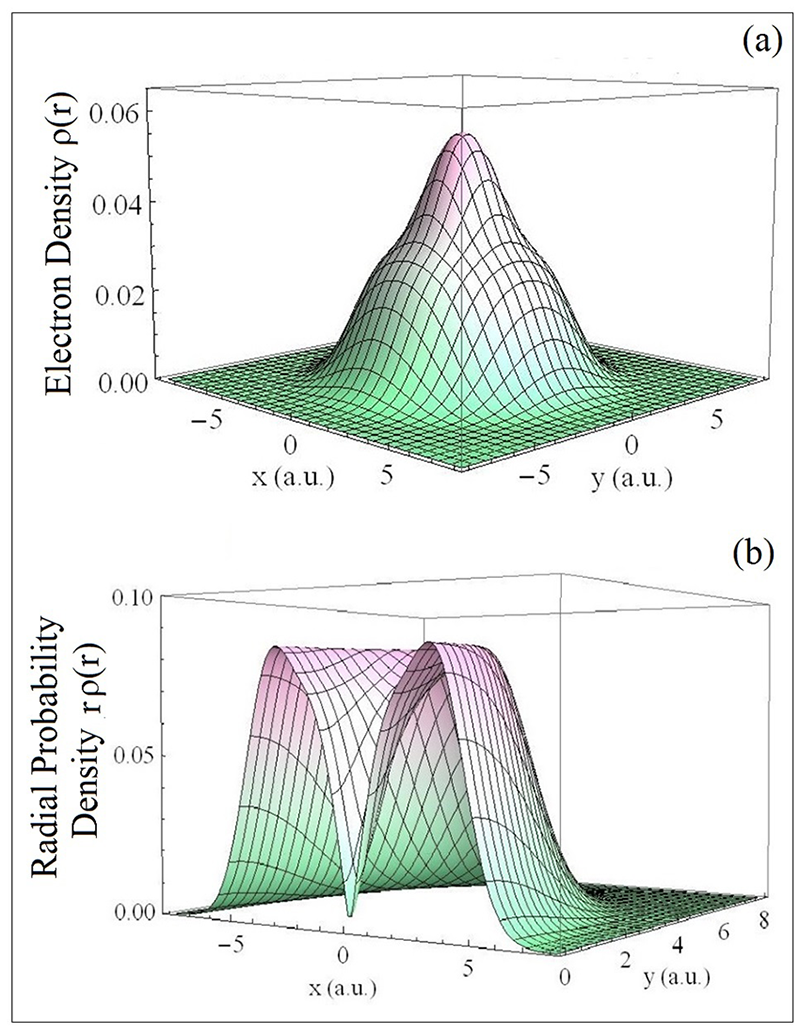}
\caption{(a) Electron density $\rho ({\bf{r}})$ of the triplet
$2^{3} S$ state of the quantum dot in a magnetic field. (b) The
radial probability density $r \rho (r)$.}
\end{figure}

\begin{figure}
\includegraphics[width=0.8\textwidth]{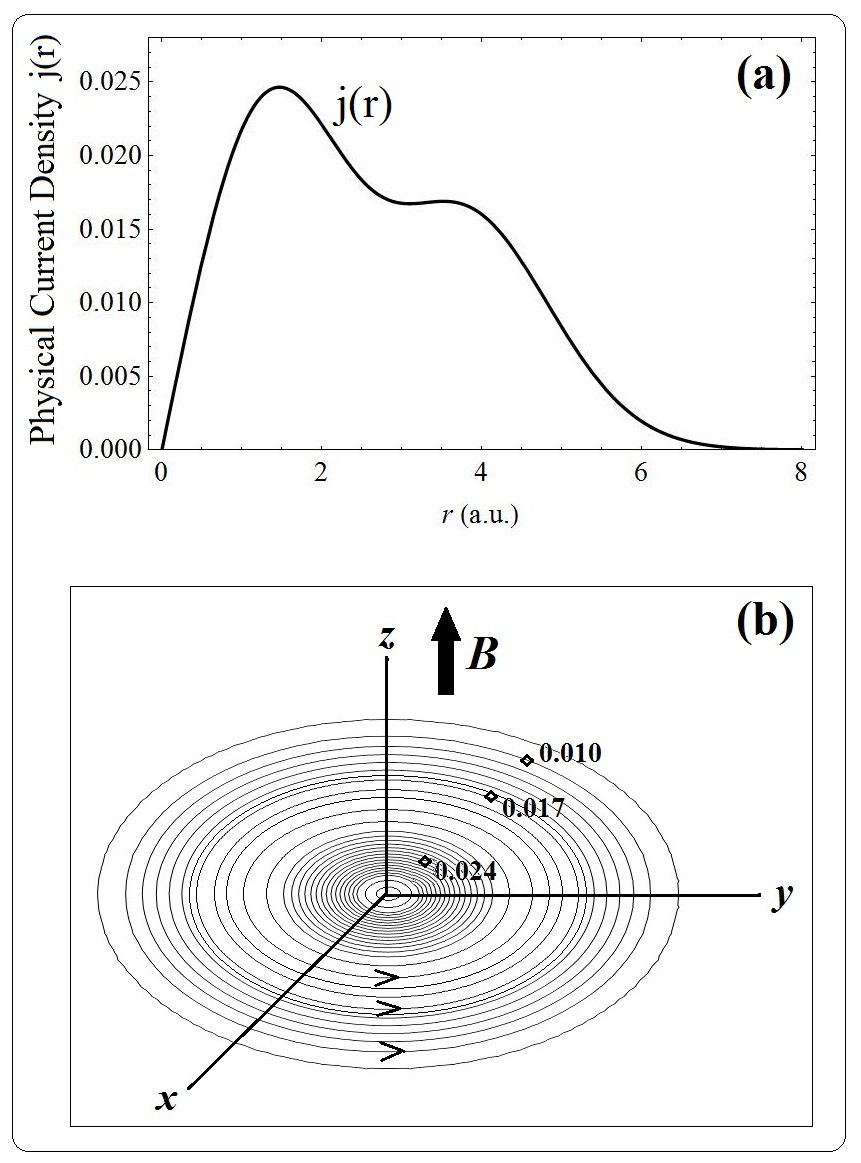}
\caption{(a) The physical current density ${\bf{j}} ({\bf{r}})$ of
the triplet $2^{3} S$ state of the quantum dot in a magnetic field
for a value of the Larmor frequency $\omega_{L} = 0.1.$ (b) The flow
line contours of ${\bf{j}} ({\bf{r}})$.}
\end{figure}

\begin{figure}
\includegraphics[width=0.8\textwidth]{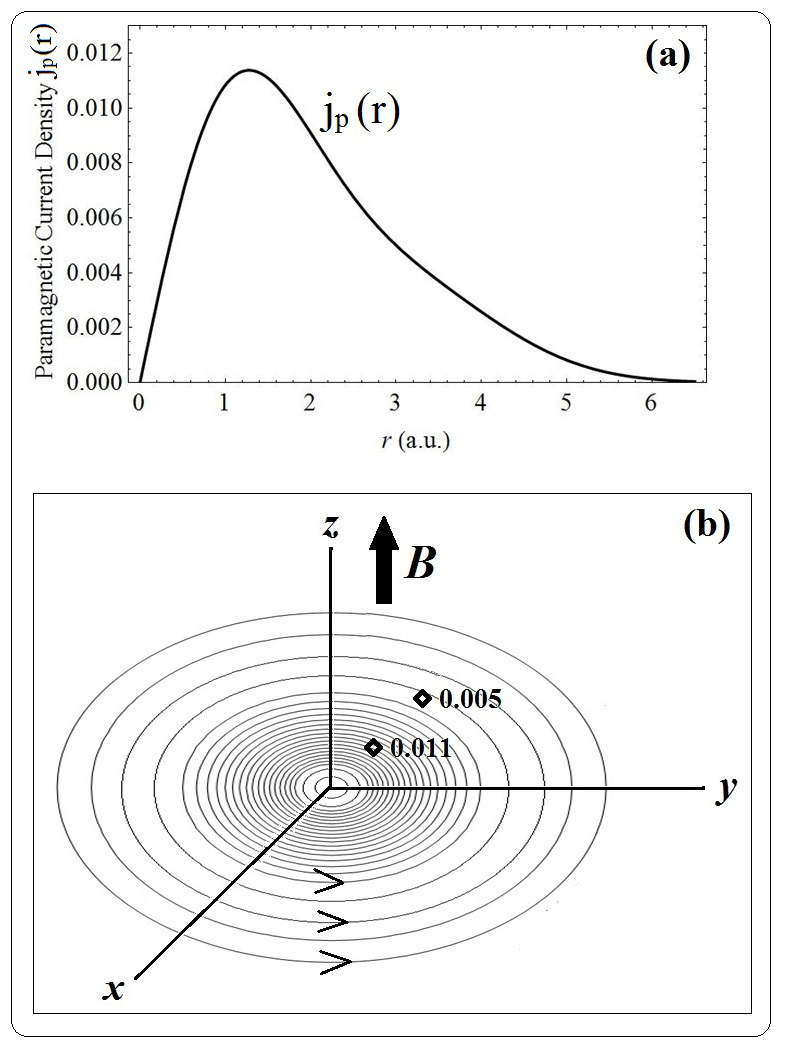}
\caption{(a) The paramagnetic current density ${\bf{j}}_{p}
({\bf{r}})$ of the triplet $2^{3} S$ state of the quantum dot in a
magnetic field. (b) The flow line contours of ${\bf{j}}_{p}
({\bf{r}})$.}
\end{figure}

\begin{figure}
\includegraphics[width=0.8\textwidth]{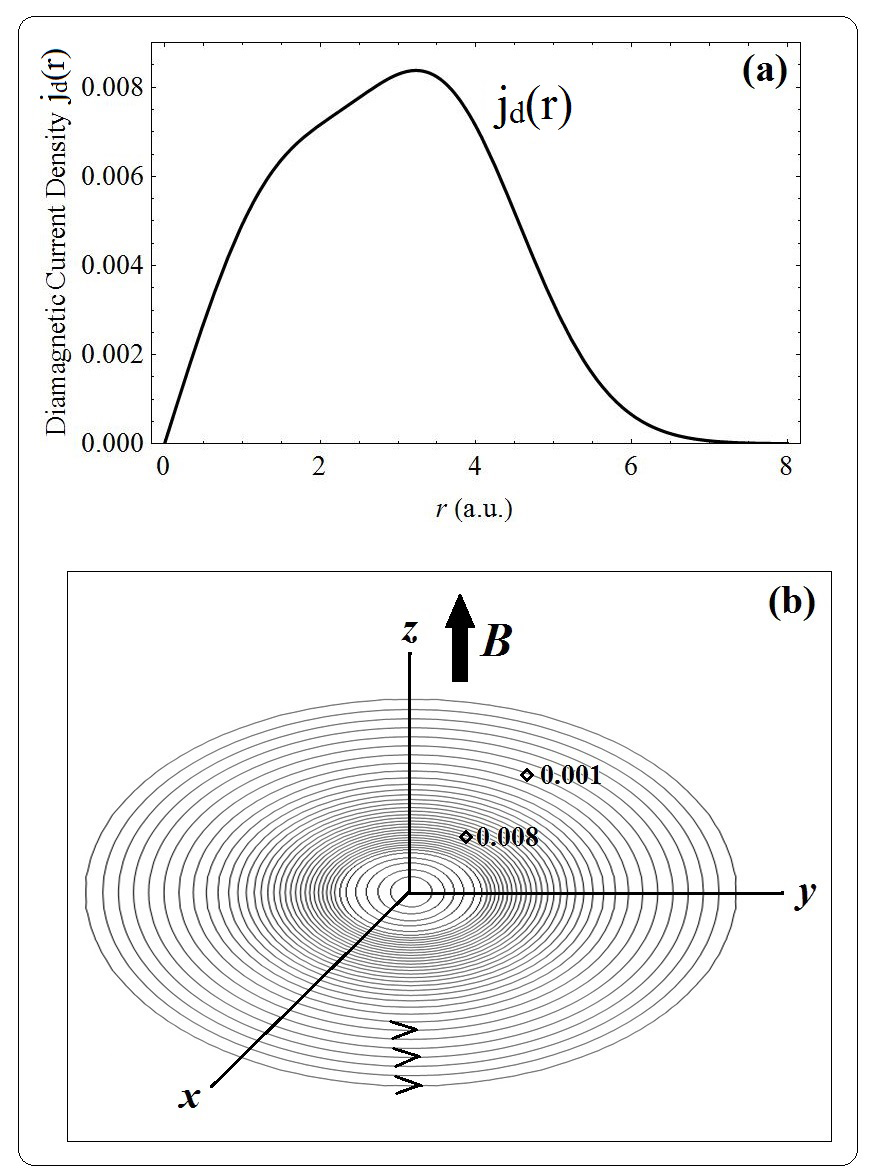}
\caption{(a) The diamagnetic current density ${\bf{j}}_{d}
({\bf{r}})$ of the triplet $2^{3} S$ state of the quantum dot in a
magnetic field for a value of the Larmor frequency $\omega_{L} =
0.1.$ (b) The flow line contours of ${\bf{j}}_{d} ({\bf{r}})$.}
\end{figure}

\begin{figure}
\includegraphics[width=0.8\textwidth]{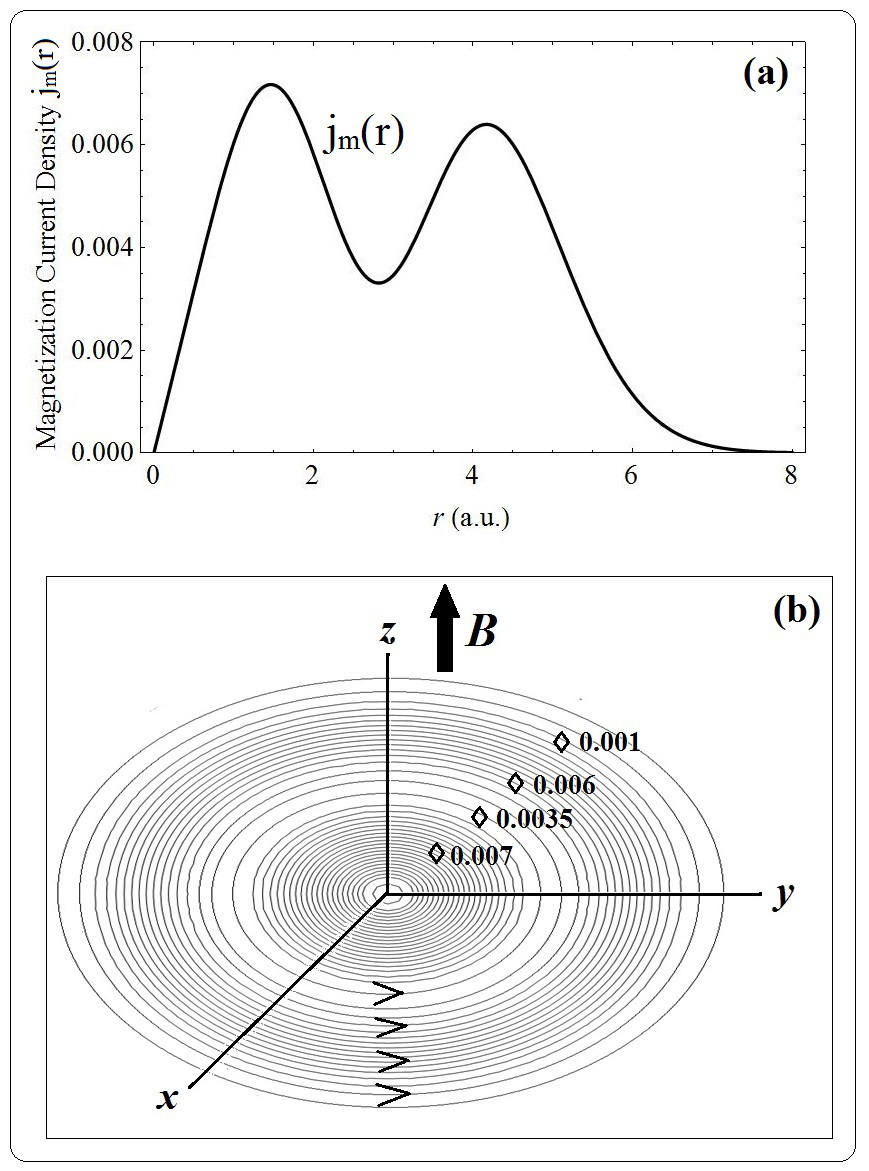}
\caption{(a) The magnetization current density ${\bf{j}}_{m}
({\bf{r}})$ of the triplet $2^{3} S$ state of the quantum dot in a
magnetic field. (b) The flow line contours of ${\bf{j}}_{m}
({\bf{r}})$. }
\end{figure}

\begin{figure}
\includegraphics[width=0.8\textwidth]{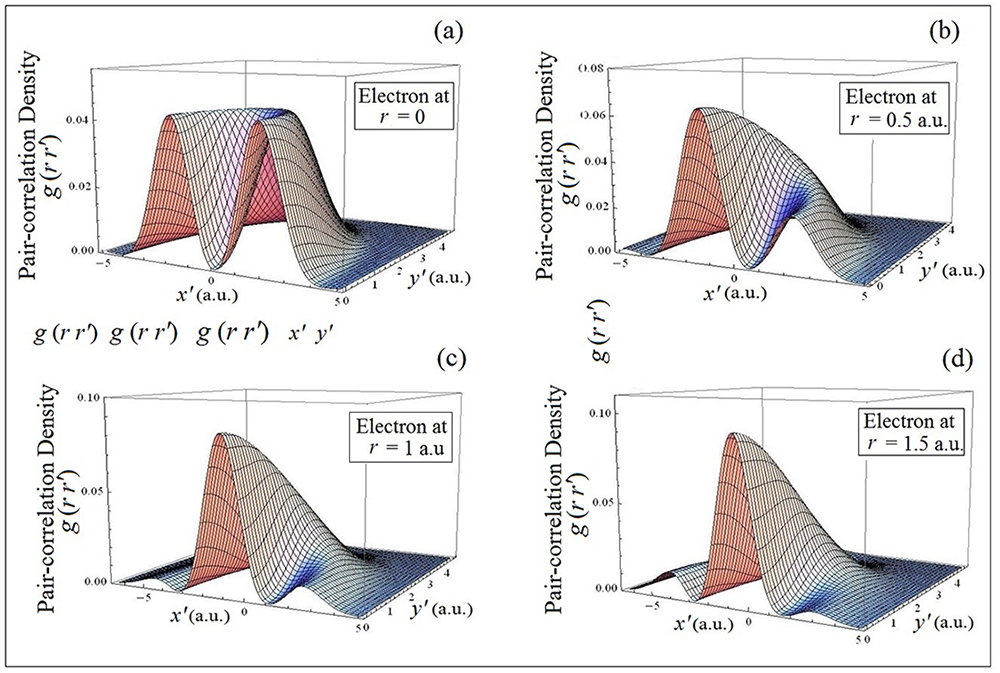}
\caption{Surface plot of the pair-correlation density $g
({\bf{rr}}')$ of the triplet $2^{3} S$ state of the quantum dot in a
magnetic field for different electron positions located on the
x-axis: (a) the center of the quantum dot at $r = 0$; (b) at $r =
0.5~a.u.$; (c) at $r = 1.0~a.u.$; (d) at $r = 1.5~a.u.$  In the
figure $x'$ is the projection of ${\bf{r}}'$ on ${\bf{r}}$,
\emph{i.e.} $x' = r' {\bf{i}}_{r} \cdot {\bf{i}}_{r'}$ , and $y'$ is
the projection of ${\bf{r}}'$ on the direction perpendicular to
${\bf{r}}$, \emph{i.e.} $y' = r'[ 1 - ({\bf{i}}_{r} \cdot
{\bf{i}}_{r'})^{2}]^\frac{1}{2}$.}
\end{figure}

\begin{figure}
\includegraphics[width=0.8\textwidth]{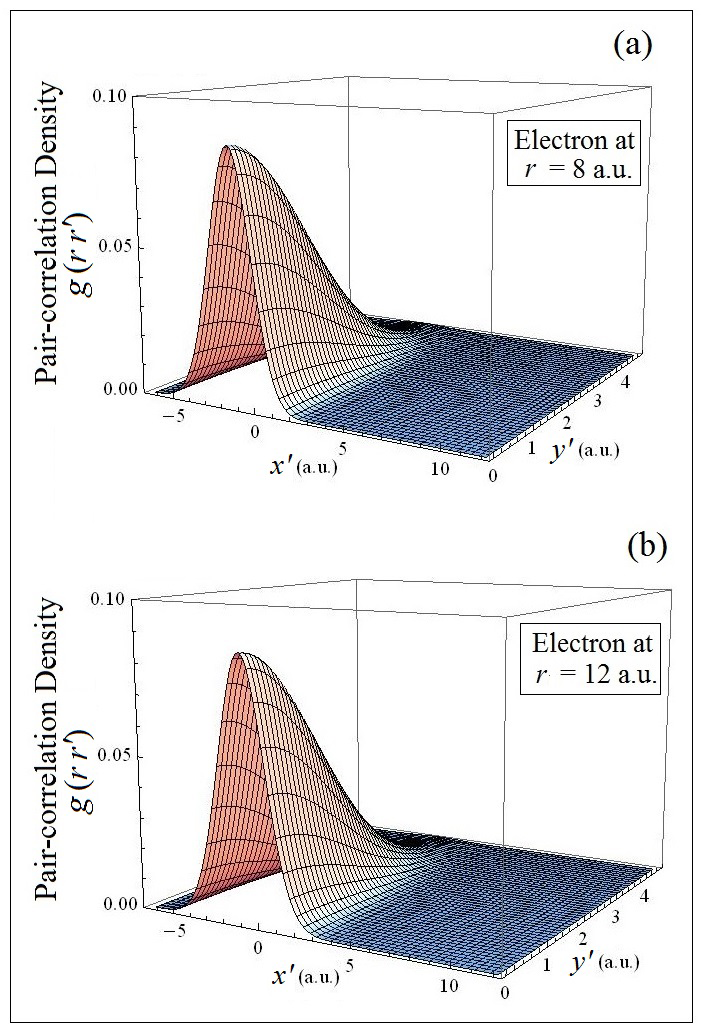}
\caption{The same as in Fig. 10 but for asymptotic electron
positions:  (a) at $r = 8.0~a.u.$; (b) at $r = 12.0~a.u.$ }
\end{figure}

\begin{figure}
\includegraphics[width=0.8\textwidth]{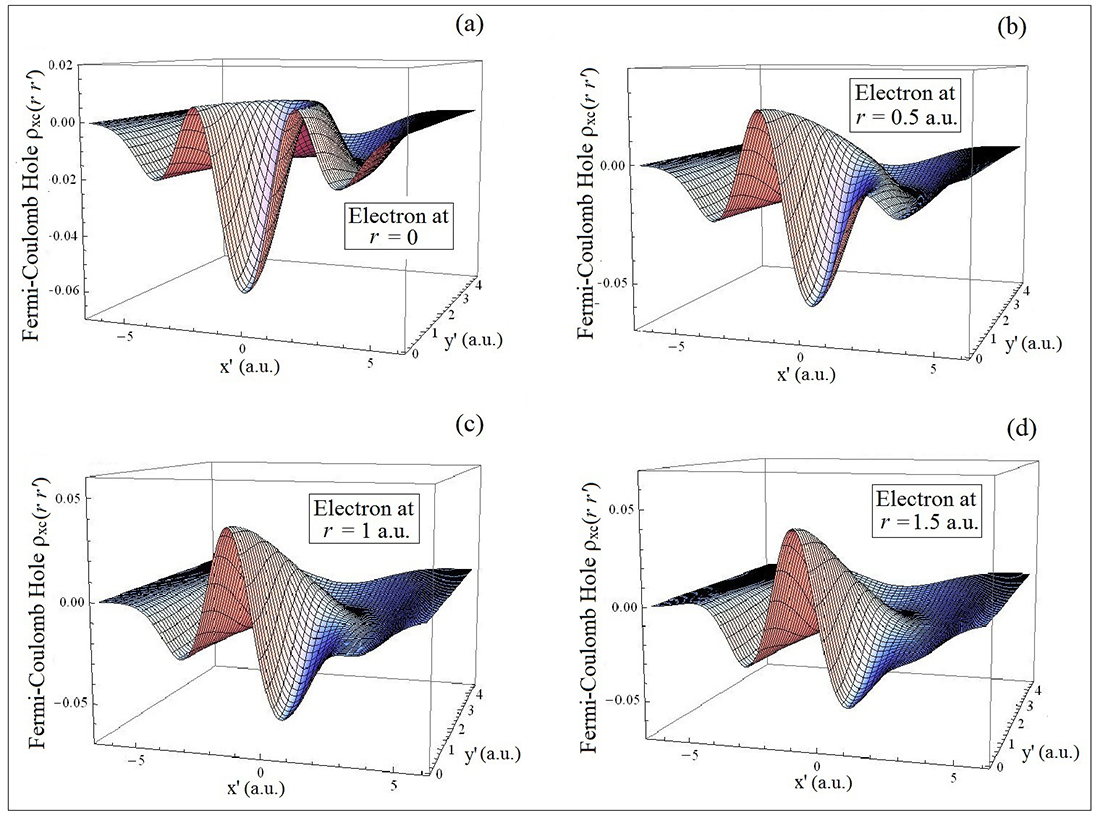}
\caption{Surface plot of the Fermi-Coulomb hole charge $\rho_{xc}
({\bf{rr}}')$ of the triplet $2^{3} S$ state of the quantum dot in a
magnetic field for different electron positions located on the
$x$-axis: (a) the center of the quantum dot at $r = 0$; (b) at $r =
0.5~a.u.$; (c) at $r = 1.0~a.u.$; (d) at $r = 1.5~a.u.$  In the
figure $x'$ is the projection of ${\bf{r}}'$ on ${\bf{r}}$,
\emph{i.e.} $x' = r' {\bf{i}}_{r} \cdot {\bf{i}}_{r'}$ , and $y'$ is
the projection of ${\bf{r}}'$ on the direction perpendicular to
${\bf{r}}$, \emph{i.e.} $y' = r'[ 1 - ({\bf{i}}_{r} -
{\bf{i}}_{r'})^{2}]^\frac{1}{2}$.}
\end{figure}

\begin{figure}
\includegraphics[width=0.8\textwidth]{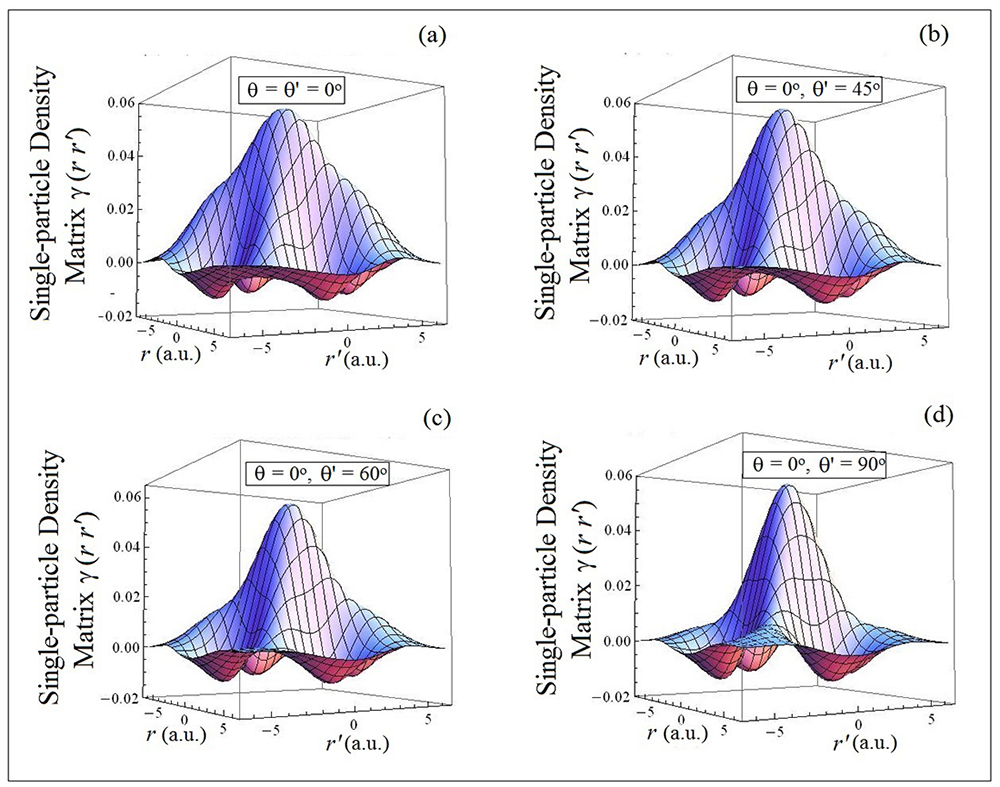}
\caption{The single particle density matrix $\gamma ({\bf{rr}}')$
for the triplet $2^{3} S$ state of the quantum dot in a magnetic
field. The panels correspond to (a) $\theta = \theta' = 0^{\circ}$;
(b) $\theta = 0^{\circ}, \theta' = 45^{\circ}; (c) \theta =
0^{\circ}, \theta' = 60^{\circ}; (d) \theta = 0^{\circ}, \theta' =
90^{\circ}.$ }
\end{figure}

\begin{figure}
\includegraphics[width=0.7\textwidth]{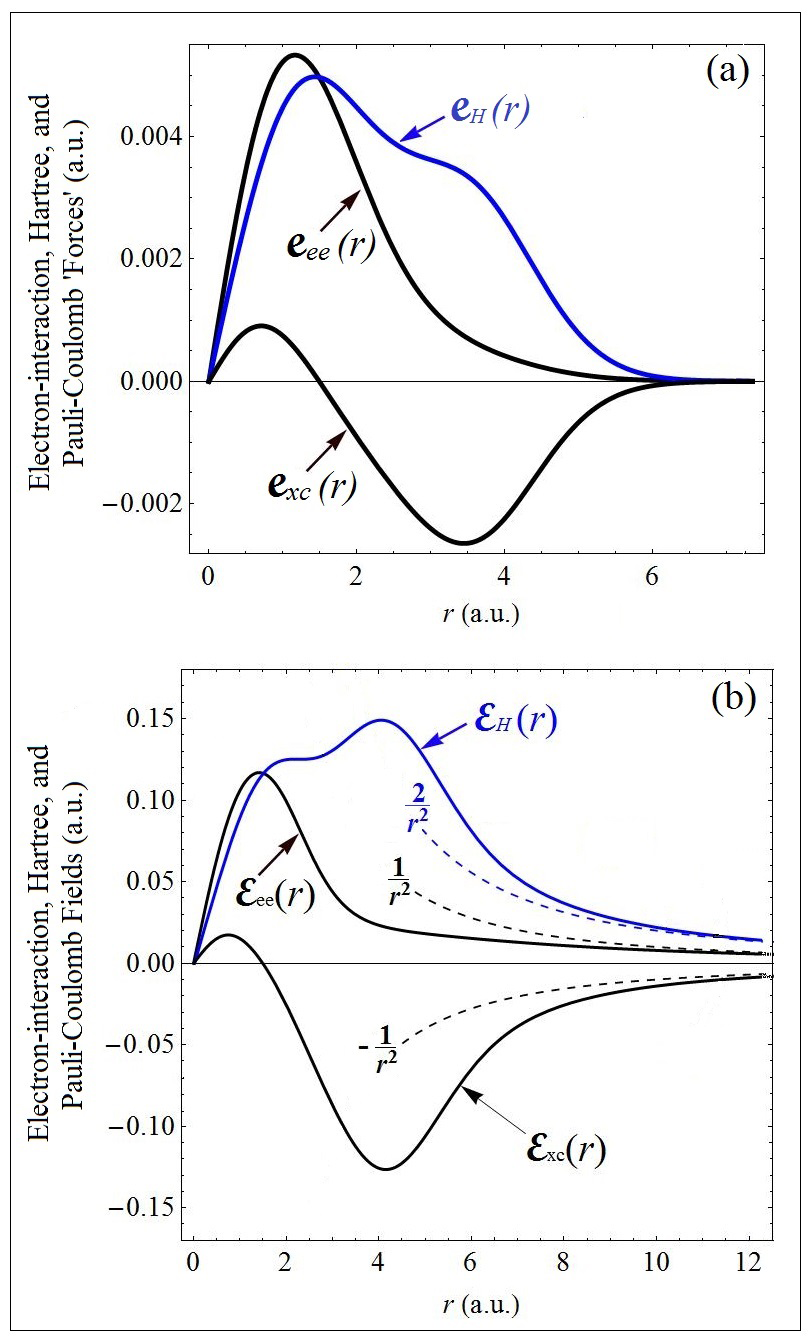}
\caption{(a) The electron-interaction ${\bf{e}}_\mathrm{ee} (r)$,
Hartree ${\bf{e}}_{H} (r)$, and Pauli-Coulomb ${\bf{e}}_\mathrm{xc}
(r)$ `forces'. (b) The electron-interaction
${\boldsymbol{\cal{E}}}_\mathrm{ee} (r)$, Hartree
${\boldsymbol{\cal{E}}}_{H} (r)$, and Pauli-Coulomb
${\boldsymbol{\cal{E}}}_\mathrm{xc} (r)$ fields. The functions
$1/r^{2}, 2/r^{2}$, and $- 1/r^{2}$ are also plotted as dashed
lines. }
\end{figure}

\begin{figure}
\includegraphics[width=0.8\textwidth]{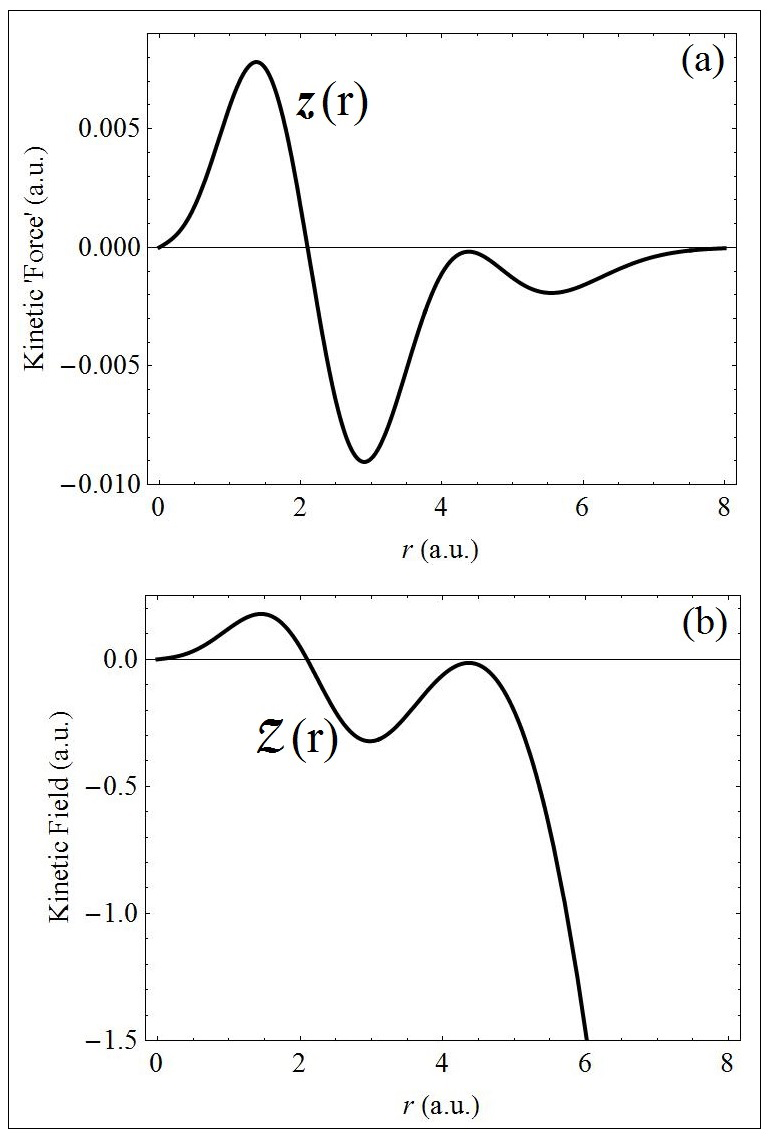}
\caption{The kinetic (a) `force' ${\bf{z}} (r)$, and (b) field
${\boldsymbol{\cal{Z}}} (r)$. }
\end{figure}

\begin{figure}
\includegraphics[width=0.8\textwidth]{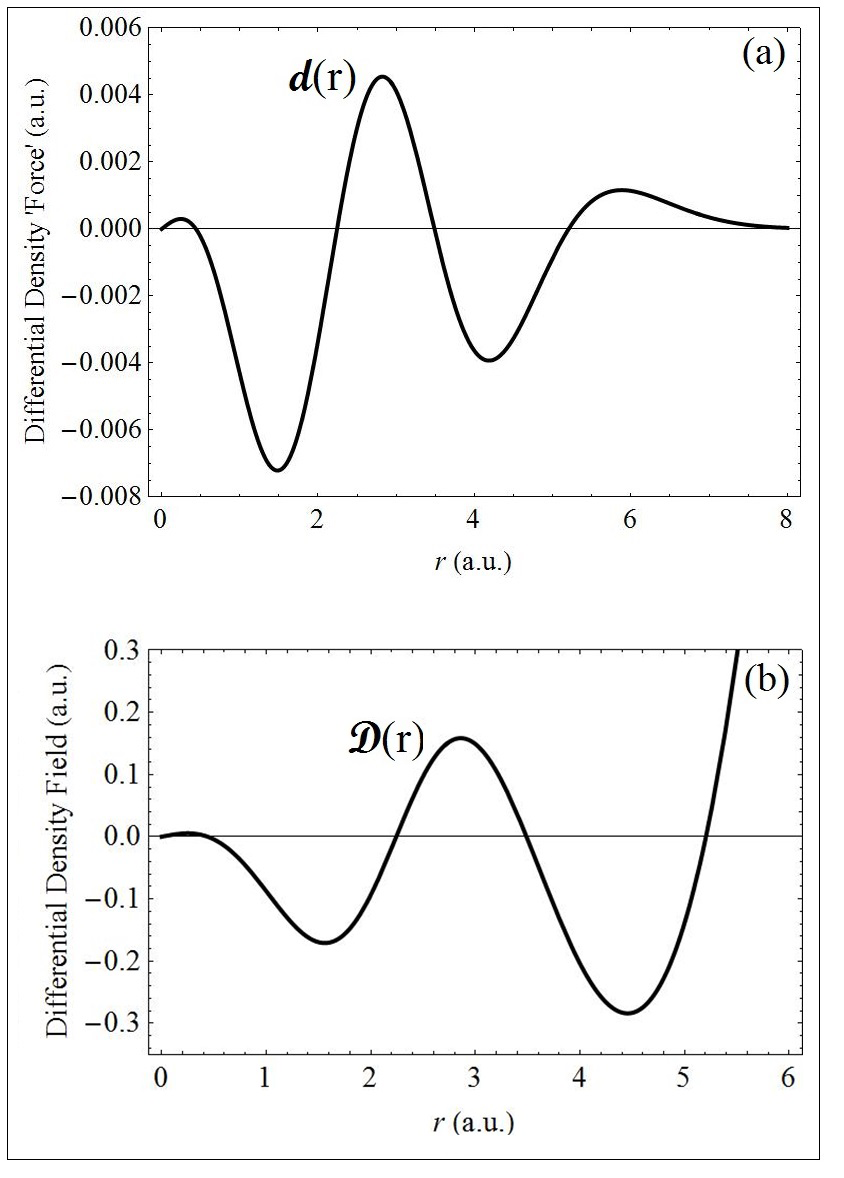}
\caption{The differential density (a) `force' ${\bf{d}} (r)$, and
(b) field ${\boldsymbol{\cal{D}}} (r)$.}
\end{figure}

\begin{figure}
\includegraphics[width=0.7\textwidth]{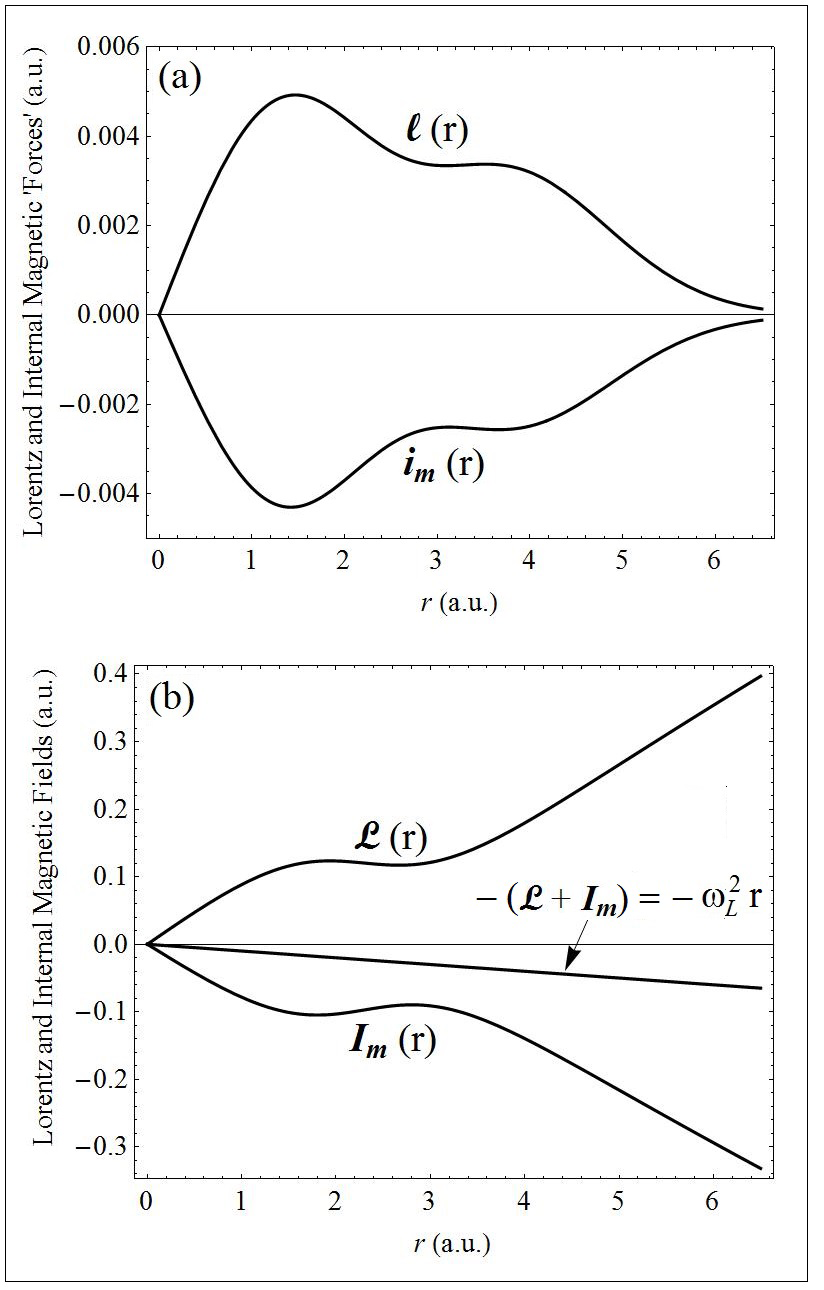}
\caption{The Lorentz and internal magnetic (a) `forces'
$({\boldsymbol{\ell}} (r), {\bf{i}}_{m} (r))$, and (b) fields
(${\boldsymbol{\cal{L}}} (r), {\boldsymbol{\cal{I}}}_{m} (r))$.  The
linear function ${\boldsymbol{\cal{M}}} (r) = -
[{\boldsymbol{\cal{L}}} (r) + {\boldsymbol{\cal{I}}}_{m} (r) ]  = -
\omega_{L}^{2} r$ is also plotted.}
\end{figure}

\begin{figure}
\includegraphics[width=0.8\textwidth]{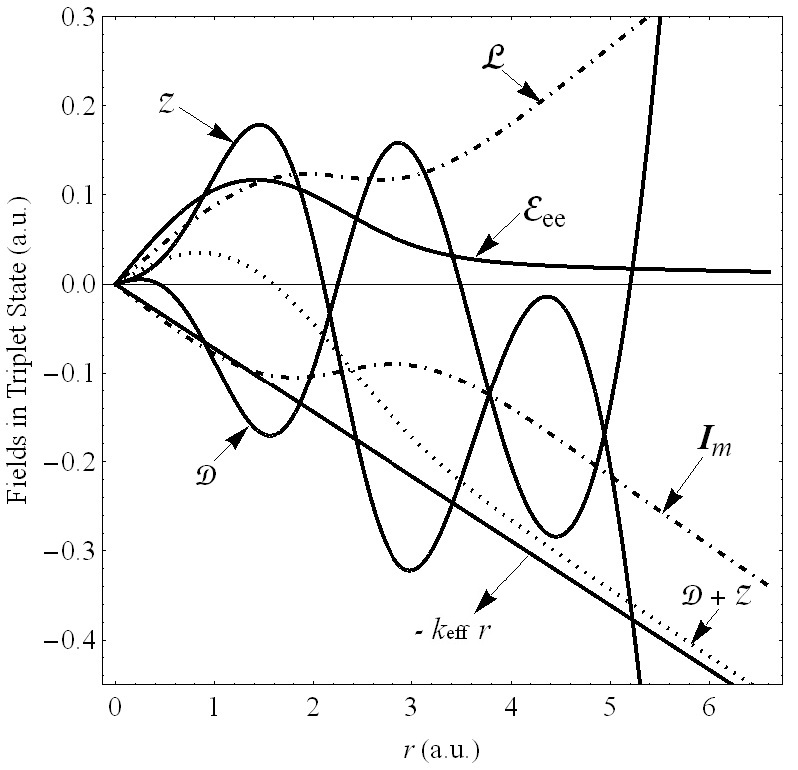}
\caption{The fields experienced by each electron:
electron-interaction ${\boldsymbol{\cal{E}}}_\mathrm{ee} (r)$;
kinetic ${\boldsymbol{\cal{Z}}} (r)$; differential density
${\boldsymbol{\cal{D}}} (r)$; Lorentz ${\boldsymbol{\cal{L}}} (r)$;
and internal magnetic ${\boldsymbol{\cal{I}}}_{m} (r)$. The fields
${\boldsymbol{\cal{L}}} (r)$ and ${\boldsymbol{\cal{I}}}_{m} (r)$
are plotted for a value of the Larmor frequency of $\omega_{L} =
0.1.$ Also plotted are the sum ${\boldsymbol{\cal{D}}} (r) +
{\boldsymbol{\cal{Z}}} (r)$, and $- k_\mathrm{eff} r$.}
\end{figure}

\end{document}